\documentclass{elsart}

\usepackage{graphicx}  

\def\ppbar{p\overline{p}}
\def\bbbar{b\overline{b}}
\def\nnbar{\nu\overline{\nu}}

\newcommand {\Dzero}      {\mbox{D{\O}}}
\newcommand{\WWWAddr}[1]     {{\tt {#1}}}

\def\insertfig#1{
  \insertxfig{#1}{5truein}
}

\def\insertyfig#1#2{
  \openin 1 #1
  \ifeof 1
    \closein 1
    \typeout{File #1 does not exist} \vskip 1truecm 
    \centerline{\fbox{FIGURE PLACEHOLDER}} \vskip 1truecm
  \else
    \closein 1
    \typeout{Reading #1}
    \centerline{\includegraphics[height=#2]{#1}}
  \fi
}

\def\insertxfig#1#2{
  \openin 1 #1
  \ifeof 1
    \closein 1
    \typeout{File #1 does not exist} \vskip 1truecm 
    \centerline{\fbox{FIGURE PLACEHOLDER}} \vskip 1truecm
  \else
    \closein 1
    \typeout{Reading #1}
    \centerline{\includegraphics[width=#2]{#1}}
  \fi
}

\begin{document}
\bibliographystyle{elsart-num}

\begin{frontmatter}

\title{The D\O\ Run II Impact Parameter Trigger 
}

\begin{flushleft}
\author{ \small
  T.~Adams$^c$,
  Q.~An$^{b,1}$,
  K.M.~Black$^{a,2}$,
  T.~Bose$^{b,3}$,
  N.J.~Buchanan$^c$,
  S.~Caron$^{d,4}$,
  D.~K.~Cho$^{a}$,
}
\author{ \small
  S.~Choi$^{i,5}$,
  A.~Das$^a$
  M.~Das$^f$,
  H.~Dong$^h$,
  W.~Earle$^a$,
  H.~Evans$^{b,6}$,
  S.N.~Fatakia$^a$,
}
\author{ \small
  L.~Feligioni$^{a,7}$,
  T.~Fitzpatrick$^e$,
  E.~Hazen$^a$,
  U.~Heintz$^a$,
  K.~Herner$^h$,
  J.D.~Hobbs$^h$,
  D.~Khatidze$^b$,
}
\author{ \small
  W.M.~Lee$^{c,8}$,
  S.L.~Linn$^{c,9}$,
  M.~Narain$^a$,
  C.~Pancake$^h$,
  N.~Parashar$^{f,10}$,
  E.~Popkov$^a$,
  H.B.~Prosper$^c$,
}
\author{ \small
  G.~Redner$^a$,
  M.~P.~Sanders$^{j,11}$,
  S.~Sengupta$^c$,
  B.~Smart$^h$,
  L.~Sonnenschein$^{a,11}$,
  G.~Steinbr\"uck$^{b,12}$,
}
\author{ \small
  W.~Taylor$^{h,13}$,
  S.~Ansermet-Tentindo$^{c,14}$,
  H.~D.~Wahl$^c$,
  T.~Wijnen$^g$,
  J.~Wittlin$^{a,15}$,
  J.~Wu$^h$,
}
\author{ \small
  S.~X.~Wu$^a$,
  A.~Zabi$^{a,16}$,
  J.~Zhu$^h$
}
\address{$^{a}$Boston University, Boston, Massachusetts 02215, USA}       
\address{$^{b}$Columbia University, New York, New York 10027, USA}        
\address{$^{c}$Florida State University, Tallahassee, Florida 32306, USA} 
\address{$^{d}$FOM-Institute NIKHEF and University of Amsterdam/NIKHEF, Amsterdam, The Netherlands}
\address{$^{e}$Fermi National Accelerator Laboratory, Batavia,            
                   Illinois 60510, USA}                                       
\address{$^{f}$Louisiana Tech University, Ruston, Louisiana 71272, USA}
\address{$^{g}$Radboud University Nijmegen, The Netherlands}
\address{$^{h}$State University of New York, Stony Brook,                 
                   New York 11794, USA}                                       
\address{$^{i}$University of California, Riverside, California 92521, USA}
\address{$^{j}$University of Manchester, Manchester, United Kingdom}

\address{$^1$Now at Dept. of Modern Physics, Univ. of Science and 
   Technology of China, Hefei, P.R. China}
\address{$^2$Now at Laboratory for Particle Physics and Cosmology, Harvard
   University, Cambridge, MA 02138, USA}
\address{$^3$Now at Brown University, Physics Dept., Providence, RI 02912, USA}
\address{$^4$Now at Physikalisches Institut, Universit\"at Freiburg,
   Freiburg, Germany}
\address{$^5$Now at SungKyunKwan University, Suwon, Korea}
\address{$^6$Now at Indiana University, Dept of Physics. Bloomington, IN 47405,
   USA}
\address{$^7$Now at CPPM, IN2P3, CNRS, Universite de la Mediterranee, Marseille, France}
\address{$^8$Now at Northern Illinois University, De Kalb, IL 60115, USA}
\address{$^9$Now at Florida International University, Miami, FL 32901, USA}
\address{$^{10}$Now at Purdue University Calumet, Hammond, Indiana 46323, USA}
\address{$^{11}$Now at LPNHE, Universites Paris VI and VII, IN2P3-CNRS, Paris, France}
\address{$^{12}$Now at Institut f\"ur Experimentalphysik, Universit\"at Hamburg, Germany}
\address{$^{13}$Now at Dept. of Physics, York Univ, Toronto, Canada}
\address{$^{14}$Now at EPFL, Lausanne, CH-1015 Switzerland}
\address{$^{15}$Now at Institute for Defense Analyses, Alexandria, VA 22311, USA}
\address{$^{16}$Now at Dept. of Physics, Imperial College London, London, UK}

\end{flushleft}

\begin{abstract}
Many physics topics to be studied by the D\O\ experiment during Run II of the
Fermilab Tevatron $\ppbar$ collider give rise to final states containing
$b$--flavored particles.  Examples include Higgs searches, top quark production
and decay studies, and full reconstruction of $B$ decays.  The sensitivity to
such modes has been significantly enhanced by the installation of a silicon
based vertex detector as part of the D\O\ detector upgrade for Run II.
Interesting events must be identified initially in 100-200 $\mu$s to be
available for later study.  This paper describes custom electronics used in the
D\O\ trigger system to provide the real--time identification of events having
tracks consistent with the decay of $b$--flavored particles.
\end{abstract}

\end{frontmatter}

Run II of the Fermilab Tevatron $\ppbar$ collider, which began in 2001, will
result in a data set 30 to 50 times larger than that from the previous run
(1992--1996), and the data are begin taken at a center--of--mass energy
$\sqrt{s}=1.96$ TeV, roughly 10\% higher than previously.  Together these
improvements will result in an unprecedented opportunity to study a variety of
interactions.  Among these are a number that result in $b$--flavored particles
in the final state including: reconstruction of $B$ decays for study of flavor
physics, $B-\overline{B}$ mixing and the resulting CKM constraints, searches
for Higgs bosons and study of the top quark.  These processes occur at a
significant rate, with hundreds (a Standard Model Higgs) to $\approx 10^{11}$
(B mesons) of signal interactions produced during Run II.  Production of
similar, non $b$--flavored background events, however, occurs at a rate orders
of magnitude higher than the signals.  At a hadron collider this poses a
particular challenge because the data from most collisions are permanently
discarded immediately following a fast and therefore rudimentary calculation of
the event properties.  The raw interaction rate is 1.7 MHz, but only 50--100
events per second can be saved for detailed study, motivating the need for
high--speed identification of interesting collisions.  The hardware and
software used for this identification are collectively called the trigger
system.

This paper describes a custom hardware trigger component, the Silicon Track
Trigger (STT), designed to identify collisions giving rise to $b$--flavored 
particles.
The first section of this paper provides an overview of the D\O\ detector and
trigger system. The second section is a functional overview of the STT.  The
third section contains a general description of the STT hardware, and the
fourth section of the paper is a detailed explanation of each major hardware
element of the STT.  The fifth section describes the STT simulator. The final
section discusses STT performance.


\section{D\O\ Detector and Trigger Overview\label{s-d0trig}} 
The D\O\ detector at the Fermilab Tevatron $\ppbar$ collider is a large,
general-purpose particle detector consisting of a magnetic spectrometer for
reconstructing particle trajectories and measuring momenta, a nearly hermetic
calorimeter for energy measurement and systems for identifying muons.  In
addition to the active detector elements, a custom data acquisition system is
used.  The data acquisition system (DAQ) provides control and synchronization,
detector read--out hardware, and, most importantly for the topic of this paper,
trigger hardware and software giving a fast, approximate selection of
interesting collisions.  The remainder of this section describes the D\O\
detector, with particular emphasis on the tracking systems that send data
to the STT and on the trigger system components needed by the STT.

\subsection{D\O\ Coordinate Systems}
Positions in D\O\ are described using a right--handed Cartesian coordinate
system with the $z$ axis along the nominal proton--beam direction, and the $x$
axis in the plane of the accelerator with positive $x$ away from the center of
the Tevatron ring. A standard cylindrical coordinate system is also sometimes
used, with the usual definitions of $r$ and $\phi$ in terms of $x$ and $y$, and
with the $z$ axes of the cylindrical and Cartesian systems the same.  Thus, an
$r\phi$ plane is also an $xy$ plane and both are perpendicular to the Tevatron
beams.

\subsection{The D\O\ Detector}
The original D\O\ detector, consisting of non-magnetic tracking, a Uranium,
liquid Argon calorimeter and muon detector, was used during Run I of
the Tevatron.  This detector configuration is described in detail
elsewhere~\cite{b-d0a}.  Significant upgrades to the tracking, muon detection
systems, data acquisition hardware and trigger systems were made between the
end of Run I and the start of Run II.  These upgrades are described in detail
in Ref.~\cite{b-d0b}.  The upgraded tracking and trigger systems provide all
data input to the STT and are described further in the remainder of this
section.

\begin{figure}
\insertfig{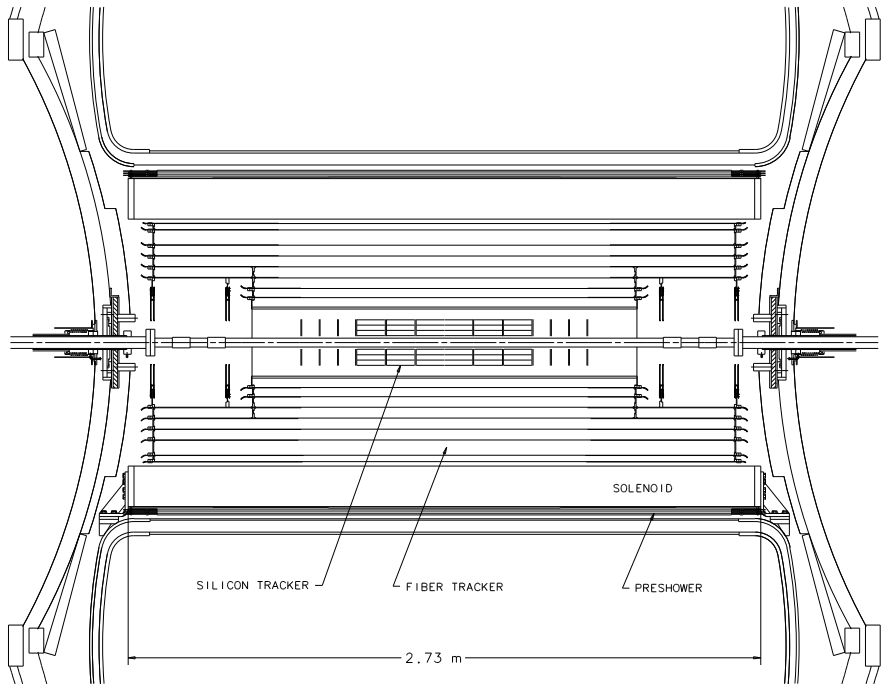}
\caption{A side view of the Run II D\O\ tracker.  This is adapted from
   Ref.~\cite{b-d0b}.
   \label{f-tracker}}
\end{figure}

The upgraded tracking system is a magnetic spectrometer whose main components
are a silicon microstrip detector (SMT) for precision vertex reconstruction,
and a scintillating--fiber based, large--volume central tracker (CFT) for
momentum measurement.  These elements are all contained inside a
superconducting solenoid producing a 2~T magnetic field.  The tracking system
is shown in Fig.~\ref{f-tracker}.  These detectors all have an approximately
cylindrical geometry with the symmetry axes of the cylinders coincident with
the Tevatron proton and anti-proton beams.

\begin{figure}
\insertfig{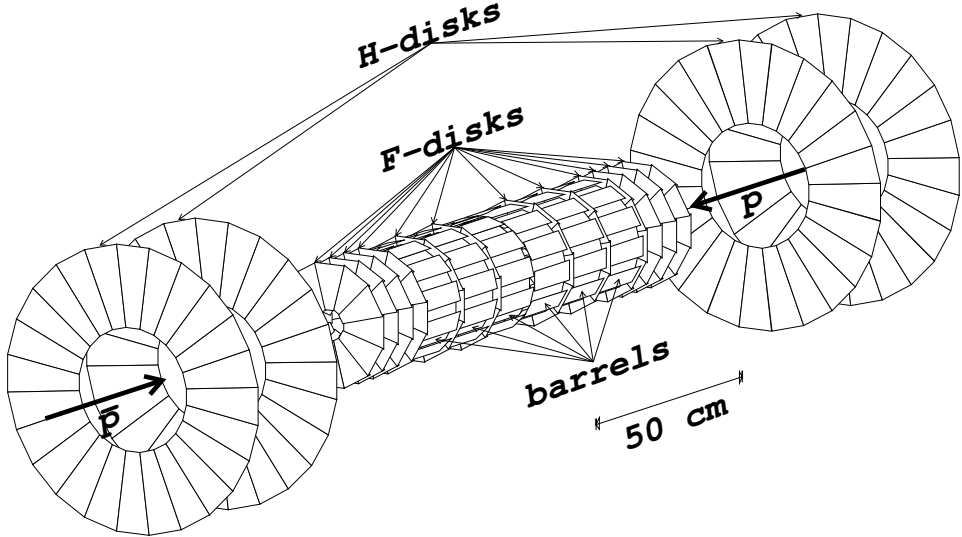}
\caption{An orthographic view of the D\O\ Silicon Microstrip Tracker (SMT).
 The barrels and disks are labeled.  Barrel data is used by the
 STT. A more detailed figure is available in Ref.~\cite{b-d0b}.\label{f-smt1}}
\end{figure}

\begin{figure}
\insertxfig{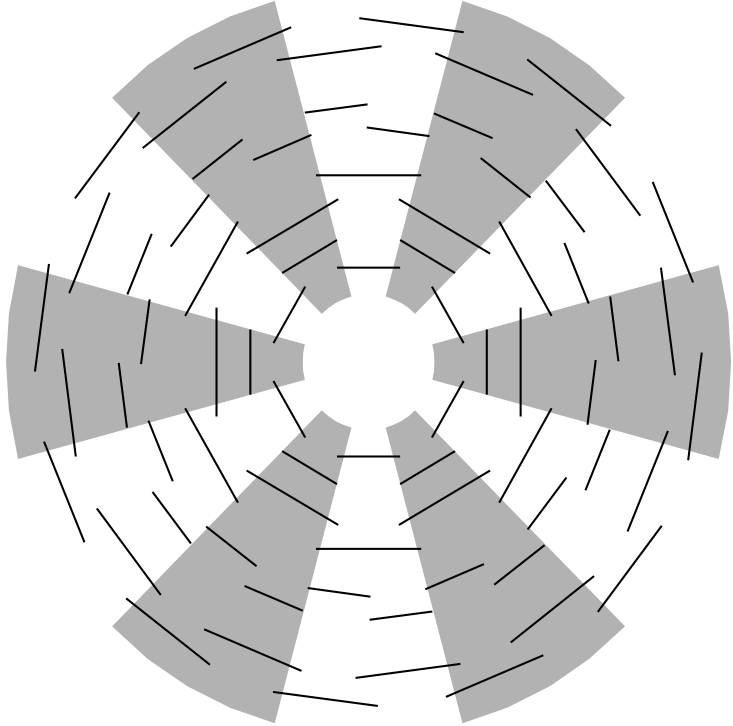}{4truein}
\caption{A simplified end view of a SMT barrel. Each line represents the projection
 of one ladder. The alternating shaded and unshaded areas show the $30^o$
 STT sectors.\label{f-smt2}}
\end{figure}

The SMT detector consists of multiple silicon--strip detectors arranged in a
barrel--and--disk geometry as shown in Fig.~\ref{f-smt1}.  The detector
provides a mixture of 3D (correlated $r\phi$ and $rz$ plane) and 2D($r\phi$ or
$rz$ plane) position reconstruction, representing an optimization between
signal--and--background discrimination and cost.  All detector elements are
used in offline track reconstruction.  However, as described below, there is
not enough time to perform a general SMT+CFT tracking algorithm in the STT, and
the initial seed tracks input to the STT have only $r\phi$ information.
Thus, only the SMT $r\phi$ information from the barrels is used by the
STT.\footnote{The SMT raw data sent to the STT have $r\phi$ and $rz$ data
interleaved, so the $rz$ positions are determined along with the $r\phi$ and
can be read out for later use.  However, this information is not used by the
STT.}  The SMT has six barrels, each consisting of four layers of overlapping
silicon--strip detector elements called ladders.  A simplified end view of one
barrel is shown in Fig.~\ref{f-smt2}.  The inner two layers have 12 ladders
each, and the outermost two layers have 24 ladders each.  The ladders in each
barrel are 12~cm long, with a small gap between barrels.  The $r\phi$ strips
providing data to the STT have a 25~$\mu$m pitch, but only every other strip is
read out, providing an effective pitch of 50~$\mu$m and intrinsic resolution of
roughly 10~$\mu$m in the measurement direction.  The SMT read--out
electronics~\cite{SVXII} have programmable threshold levels, and only detector
strips with pulse heights higher than threshold are read out.  The read out
time thus depends on the hit--strip multiplicity.  There is a 5~$\mu$s latency
before read out begins, and the read--out typically adds an additional 
10~$\mu$s.

The CFT detector consists of multiple cylindrical shells (layers) of
scintillating fibers.  Two basic fiber arrangements are used throughout the
CFT: (1) an axial geometry with the fibers in a layer parallel to the center
axis of the cylinder providing measurement in the $r\phi$ plane, and (2) a
small--angle stereo geometry in which the fibers on the cylindrical surface are
canted at 2$^o$ with respect to the direction of axial fibers.  A pair of axial
layers and a pair of stereo layers are mounted to a common carbon--fiber
support tube to form a super-layer. The two axial fiber layers in a super-layer
have a half--fiber relative offset in $\phi$.  The same is true for the stereo
fibers. There are eight superlayers in the complete CFT, at radii of
approximately 20~cm, 25~cm, 30~cm, 35~cm, 40~cm, 45~cm, 49~cm and 52~cm.
Fibers in the inner 2 superlayers are roughly 1.7~m long in $z$; the remainder
are 2.5 m.  The individual fibers are 835~$\mu$m in diameter, and the $r\phi$
position resolution is roughly 250~$\mu$m.

\subsection{The D\O\ Trigger System}
\begin{figure}
\insertfig{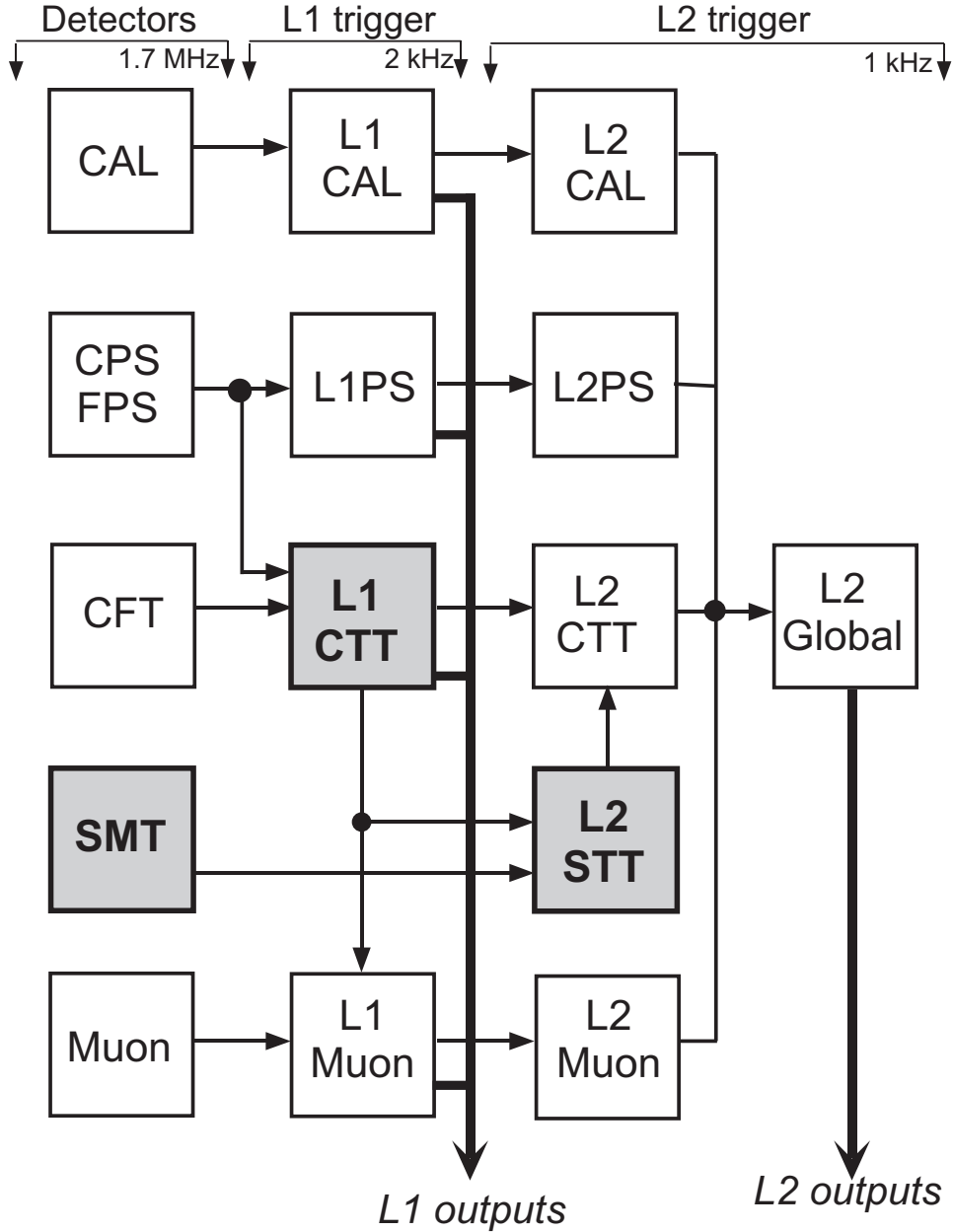}
\caption{A block diagram of the Level 1 and Level 2 trigger system.
  \label{f-trigarch}}
\end{figure}

As previously mentioned, the raw interaction rate at the Tevatron is 1.7~MHz, but
the D\O\ rate to tape is limited to 50-100~Hz.  A highly parallel, multi-level
trigger system is used to provide the necessary rate reduction while keeping
the data for the interesting but rare interactions.  The first level
trigger (L1) consists of fast, dedicated hardware which finds energy in the
calorimeter and matches data with precoded patterns in the CFT, preshower and
muon systems.  Limited spatial correlations can also be made.  The second
level (L2) consists of custom preprocessor elements running specialized software
algorithms.  The L2 preprocessor results are fed to a L2 global processor, which
is used to make all L2 event reject or accept decisions.  The overall
architecture of L1 and L2 is shown in Fig.~\ref{f-trigarch}.  The third
level (L3) consists of a farm of commercial processors running a simplified
version of the offline detailed reconstruction code.

The L1 input event rate is 1.7~MHz, and the output rate is 2~kHz.
This is sent to L2, which must reduce this to 1~kHz.  Finally, L3 reduces the
1~kHz rate to the final 50-100~Hz rate.  The L1 trigger elements must produce
results within 3.5~$\mu$s~\cite{b-d0b}.  Queuing studies using simulated events
indicate that a L2 preprocessor has roughly $50\ \mu$s/event for processing,
and that the L2 global processor has an additional $50\
\mu$s. Up to $100\ \mu$s of latency (typically introduced by internal data 
transfer and buffering times) is also allowed.

As can be seen in Fig.~\ref{f-trigarch}, each detector element has corresponding
trigger hardware.  The STT is a L2 preprocessor, and it receives trigger data
from the L1 central track trigger (CTT) and zero--suppressed raw data from the
SMT detector.  The figure does not show the control and synchronization signals
from the trigger system used by the STT.  The read out time of the SMT data
and the use of CTT results from L1 prohibit the STT from
functioning as an L1 trigger element.

The CTT data are used as input to the STT.  The CTT L1 processor uses axial
fiber data from the CFT detector to reconstruct particle trajectories (tracks)
in the $r\phi$ plane.  This system is described in detail
elsewhere~\cite{b-d0b}.  The output sent to the STT by each of six 80$^o$
overlapping CTT $\phi$ regions consists of a header record, a list of tracks
found by the CTT and a trailer record.  No more than 46 tracks can be sent from
each CTT region for a total bound of 276 tracks/event.  The information for a
single CTT track consists of the axial fiber number in the outermost CFT layer,
the track momentum\footnote{The momentum is given as a value in one of four
ranges, $1.5\le p_T \le 3$~GeV/$c$, $3\le p_T\le 5$~GeV/$c$, $5\le p_T \le
10$~GeV/$c$ and $p_T>10$~GeV/$c$.} in the $r\phi$ plane ($p_T$) and encoded
information that allows the track position at the innermost CFT layer to be
determined.  When the trajectory and momentum are calculated by the CTT, it is
assumed that the particle has zero lifetime and that the interaction creating
the particle occurred at the center of the beam.  These assumptions allow
nearly 100\% efficiency for reconstructing particles satisfying
$p_T>1.5$~GeV/$c$ which are created within 1~mm of the average interaction
point in the $x--y$ plane.  There is negligible efficiency for particles that
do not satisfy either of these requirements.

The known positions at the innermost (A--layer) and outermost (H--layer) CFT
layers and the assumption that the particle comes from the average beam
position, determined on a run-by-run basis, provide three points needed to
define an approximate circular trajectory.  Because the STT is used to select
events in which particles travel a short distance, typically less than
500~$\mu$m before they decay, the STT track reconstruction removes the
assumption that the particle comes from the average beam position in the final,
detailed phases of the calculations.

\section{Silicon Track Trigger: Functional Overview\label{s-stt}} 


Most of the detected particles resulting from a $\ppbar$ collision emanate from
the collision point. When their trajectories are reconstructed, these particles
thus appear to come from a common production point called the primary
vertex. In contrast, $b$-flavored hadrons have a lifetime of roughly 1.5~ps,
and for typical events with $B$--hadrons produced in $\ppbar$ collusions at
$\sqrt{s}=1.96$~TeV, they travel about 0.5--1~mm before
decaying. The origins of the trajectories of particles resulting from the
$b$--quark decay thus do not originate at the primary vertex.  The distance
from the primary vertex to the point on the trajectory closest to the primary
vertex is called the impact parameter, denoted $b$.  Thus, tracks with
significantly non-zero impact parameters are a hallmark of the decay of
particles containing a $b$--quark.




The STT is designed to perform a real time calculation of the impact parameter
for trajectories reconstructed in the $r\phi$ plane.  A particle trajectory,
and thus the impact parameter, is determined by fitting an approximate circular
trajectory to points along the trajectory measured by the D\O\ SMT and CFT.
The determination uses either three or four points (clusters) measured in the
SMT and two additional points inferred from the CFT detector.  The
reconstructed impact parameters are then used to select long-lived particles.
A functionally similar system is also used by the CDF experiment, but the
design details differ considerably.~\cite{b-cdfsvt}

Although the true impact parameter of particles produced at the primary vertex
is identically zero, finite measurement resolution causes these particles to be
reconstructed with a small, but non-zero impact parameter.  The impact
parameter distribution for these particles is a Gaussian with zero mean.  The
impact parameter resolution\footnote{See section~\ref{s-performance} for
details.} for STT tracks is
\begin{equation}
\sigma_b = \sqrt{19^2 + \left(\frac{54\ \mathrm{GeV/c}}{p_T}\right)^2}\ 
  \mu\mathrm{m},
\end{equation}
with $p_T$ being the transverse momentum of the particle in GeV/$c$. The
momentum dependence arises from multiple Coulomb scattering in the beam pipe
and the SMT detector.

The Tevatron beam has a circular cross-section in the $xy$-plane with a
Gaussian width of $28--35\ \mu$m depending on accelerator settings and
instantaneous luminosity.  This causes the primary vertex to be distributed
about the beam center.  Because the production point is not determined on an
event-by-event basis in the L1 or L2 trigger, this increases the apparent
impact parameter resolution resulting in a final total impact parameter
resolution of
\begin{equation}
  \sigma_b = \sqrt{35^2 + 19^2 + \left(\frac{54\ \mathrm{GeV/c}}{p_T}\right)^2}
    \ \mu\mathrm{m} 
     = \sqrt{40^2 + \left(\frac{54\ \mathrm{GeV/c}}{p_T}\right)^2}\ \mu\mathrm{m}.
\end{equation}
again with $p_T$ in units of GeV/$c$.
One physics topology likely to benefit from the STT is the Higgs search using 
the final state $\ppbar\rightarrow ZH\rightarrow (\nu\overline{\nu})(\bbbar)$
For this mode, the average impact parameter for particles arising from
$B$--hadron decay is roughly 80 $\mu$m, so considerable sensitivity remains.

The situation is further complicated because the Tevatron beams are not
necessarily centered exactly on the SMT $z$-axis, nor are the beams exactly
parallel to this axis.  The position and tilt of the beams are measured by D\O\
and transmitted to the STT roughly once every four hours, and the STT
computation includes correction factors such that reported impact parameters
are calculated with respect to the measured beam position.  Residual effects
from beam tilt increase the resolution by less than 5~$\mu$m from that quoted
above.  Including these effects reproduces the observed impact parameter (IP)
width for 
STT tracks from collider data.

The STT hardware provides the following functionality: \newline
\begin{enumerate}
 \item{Input from and synchronization with the D\O\ trigger and the 
     Tevatron,}
 \item{Receiving initial tracks from the CTT,}
 \item{Receiving raw data from the SMT,}
 \item{Performing position reconstruction using SMT raw data,}
 \item{Associating SMT clusters with input CTT tracks,}
 \item{Selecting the subset of associated clusters used 
   in fitting,}
 \item{Fitting the SMT and CTT information to
     determine the impact parameter $b$, }
 \item{Writing CTT data and STT
     results to the D\O\ L2 Central Track Trigger (L2CTT) and}
 \item{Storing data for read out to L3.}
\end{enumerate}
The processing is performed in the above order.  However, significant parallel
capabilities and pipelines in the hardware permit data from different parts of
the detector and from different interactions to be processed simultaneously.
In addition to the impact parameter, the fit in step (7) also determines the
track direction and momentum in the $r\phi$ plane.  The STT hardware
architecture and custom electronics are described in overview in
section~\ref{s-stt-hardware} and in detail in section~\ref{s-stt-detail}.

The remainder of this section describes the inputs to the STT, the overall
synchronization with the D\O\ trigger and the Tevatron, the approach used to
determine the impact parameter for each track including cluster finding,
pattern recognition and track reconstruction, the STT output data and other
aspects of the STT system.

\subsection{Synchronization and Input Data\label{ss-go}}
Overall synchronization with the accelerator and the D\O\ trigger and detector
read out is provided by experiment-wide control logic external to the STT.  The
STT receives trigger and beam crossing information via D\O\ control
hardware called the serial command link (SCL)~\cite{SCL}.  Processing is
performed by the STT only for collisions initially selected by an L1
trigger. The STT requires two types of input data:
\begin{enumerate}
  \item The full set of tracks found by the CTT, and 
  \item The sparsified, i.e. strips with signals below a threshold are
  suppressed, raw data from the SMT barrel detectors.
\end{enumerate}
The SCL and CTT data are received and rebroadcast within the STT by the fiber
road card (FRC) described in section~\ref{ss-frc}.  The SMT raw data is
received and processed by the silicon track card (STC) described in
section~\ref{ss-stc}.


\subsection{Clustering Algorithm\label{ss-clustering}}
\begin{figure}
\insertfig{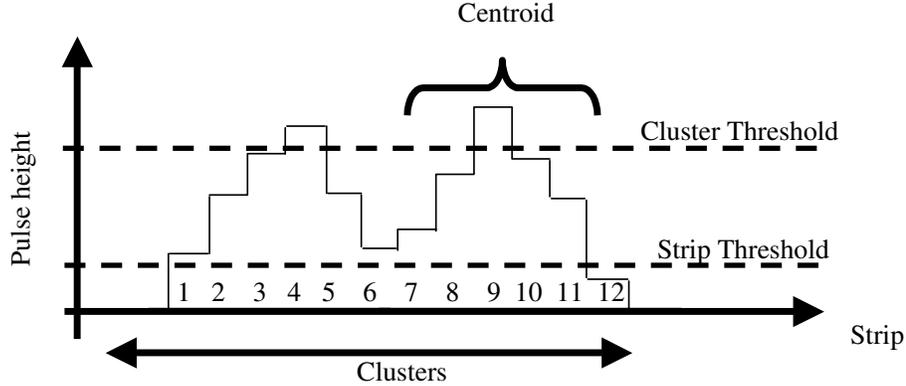}
\caption{An illustration of the SMT clustering algorithm used by the STT.
  Each bin represents the pulse height on one SMT strip, and the strips are
  adjacent in the detector.\label{f-clus}}
\end{figure}

The STT clustering algorithm is used to determine the position in the SMT at
which a charged particle passed through an SMT ladder. The charge liberated by
a single particle as it traverses an SMT ladder is typically collected by a
small number of adjacent SMT read out strips.  To obtain optimal position
resolution, this effect must be considered in the position determination.
Therefore, adjacent strips with pulse heights above a strip threshold are
grouped into clusters and for each cluster a pulse-height weighted centroid is
calculated.  Fig.~\ref{f-clus} illustrates the clustering algorithm.

The cluster-finding algorithm works in the following manner. For each ladder,
all axial strips are scanned in order. A cluster is a sequence of neighboring
strips, each with a pulse height above the strip threshold. A centroid is only
calculated if at least one strip in the cluster has a pulse height above the
centroid threshold. The centroid position is calculated based on five strips
centered on the strip with the largest pulse height.\footnote{An option is also
available to compute the centroid based on three strips.  However, this has not
been used during running.}  Strips outside this window are ignored in the
centroid calculation.

First the centroid is computed relative to the center strip: 
\begin{equation}
\Delta x = {-2q_1 -q_2 +q_4 +2q_5\over q_1+q_2+q_3+q_4+q_5},
\end{equation}
where the $q_i$ are the pedestal and gain corrected pulse heights of the five
strips.\footnote{The pedestal and gain information is read from the D\O\
offline database. It is updated as needed.} This relative centroid $\Delta x$
is converted to the binary value $d$, with
\begin{equation}
d=\left\{
\begin{array}{c@{\rm\ if\ }r@{|\Delta x|}l}
000 &       &\leq0.125 \\
001 & 0.125<&\leq0.375 \\
010 & 0.375<&\leq0.625 \\
011 & 0.625<&\leq0.875 \\
100 & 0.875<&          \\
\end{array}\right.
\end{equation}
Then the centroid $x$ is calculated to quarter strip precision as
\begin{equation}
x=\left\{
\begin{array}{cl} 
c\times4-d & \mbox{if $\Delta x<0$}\\
c\times4+d & \mbox{else}
\end{array}
\right.,
\end{equation}
where $c$ is the 11-bit address of the center strip. Thus the centroid $x$ is a
13-bit number.

The clustering is performed by the STC hardware described
in section~\ref{ss-stc}.  A transformation from detector-based
coordinates to standard D\O\ $(r,\phi)$ coordinates is made prior to the final
cluster filtering in the track fitting card (TFC) described in 
section~\ref{ss-tfc}.

\subsection{Pattern Recognition\label{ss-patrec}}
Constraints imposed by the L2 time budget leave too little time for a general
pattern recognition algorithm to be performed, so a simplified algorithm is
used by the STT.
\begin{figure}
\insertfig{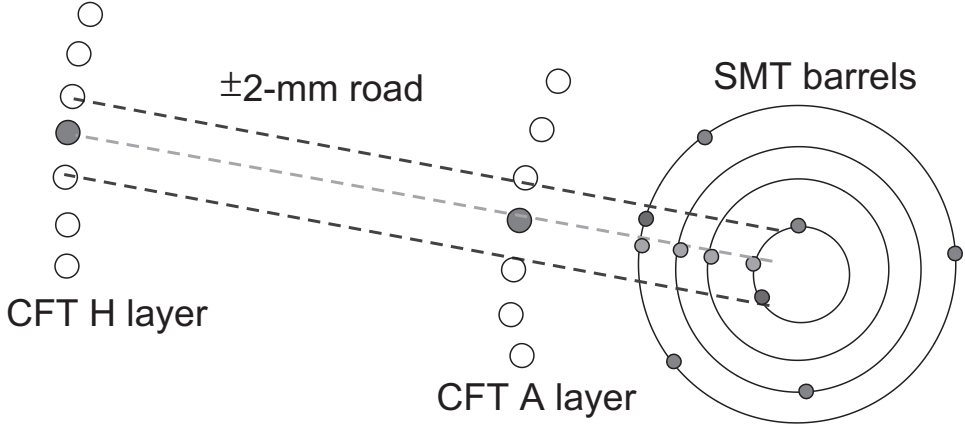}
\caption{The pattern recognition algorithm used by the STT. A $\pm 2$~mm 
 region in the $xy$ plane  (called a road)
 centered on a CTT track is initially searched for
 reconstructed SMT clusters.  A second step refines the selection, choosing a
 single cluster from each SMT layer to be used in the final track
 fitting. \label{f-road}}
\end{figure}

The first phase of the STT pattern recognition, shown in Fig.~\ref{f-road},
uses a $\pm2$~mm region (called a road) centered on each CTT input track as an
initial estimate of a possible STT track.  All $r\phi$ clusters in an event are
checked to see if they lie within any road in the event.  Each cluster that
meets this criterion is then associated with every road it overlaps.  This is
performed for all roads in the event, resulting in lists of SMT cluster to CTT
track associations.  The road width of $\pm2$~mm was chosen to maintain 
high acceptance for tracks from $B$--hadron decays and to reduce the
impact of run-to-run beam spot shifts.  This provides greater than 98\%
efficiency for SMT cluster association for any particle having 
$p_T>1.5$~GeV/$c$ and $|b|<2$~mm which is in the SMT geometric acceptance and
which passes through the points in the CFT A-layer and H-layer defined by the
CTT track.

The first phase of the pattern recognition typically results in more than one
SMT cluster per CTT track per SMT layer as described in
section~\ref{s-performance}.  The true particle trajectory passes through a
given SMT layer at only one point.  Additional clusters in a layer arise from a
variety of effects including $\delta$-ray production, clusters from nearby
particles in the event and electronics noise.  The fitting can use no more than
one cluster per layer, so a second pattern-recognition pass (called final
filtering) is made following the initial cluster to road association.  This
phase selects a single cluster on each layer for use in the fitting.

In the first step of the final filtering, the road width is narrowed to
$\pm1$~mm, and initially selected clusters outside this road are
discarded.\footnote{ The $\pm1$~mm road width can be used in the initial
selection.  However, the hardware used in the initial selection must then be
reconfigured when the beam spot moves by a modest amount.  This computation is
time consuming, so the roads are left wider, allowing more beam spot motion
before a reconfiguration is required.  The beam spot computations used in the
final selection and fitting introduce a negligible overhead.  The
reconfiguration for these is performed at the start of each run, roughly once
every four hours.} At the same time, the limited $z$ information available is
used to further filter clusters.  Because tracks are approximately straight
lines in the $rz$ plane, the pattern of SMT barrels that can contribute
clusters to the tracks is constrained.  Clusters retained for later
consideration thus are required to form a trajectory consistent with a straight
line in the $rz$-plane.  Specifically, all clusters selected by the final
filtering must be from the same SMT barrel, or from two adjacent barrels with
at most one transition from one barrel to the other.  The starting barrel is
defined as the barrel of the cluster selected by the final filtering in the
outermost silicon layer.  Only clusters that  meet this criterion are
saved for future use after the first step.

For the second step of the final filtering, the $r\phi$ distance between the
CTT track and each remaining associated cluster is calculated.  The distance
$r_i\delta\phi$ is given by
\begin{equation}
  r_i\delta\phi = |r_i(\phi_i-\phi_{0_{CTT}}) - \kappa_{CTT}\cdot r_i^2 - b_{CTT}|,
\end{equation}
in which $(r_i,\phi_i)$ is the position of a cluster associated with the track
and $\phi_{0_{CTT}}$ and $\kappa_{CTT}$ are the azimuthal angle of the track at
the measured beam position and its curvature respectively.  $b_{CTT}$ is the
impact parameter of the track relative to the nominal D\O\ coordinate origin,
computed assuming the track comes from the center of the beam.  For each SMT
layer, the cluster closest to the CTT tracks is used in determining the final
trajectory.

CTT tracks are assumed to come from the beam center.  This assumption could
bias the final filtering in favor of clusters giving fitted tracks with small
impact parameters.  A number of more flexible algorithms were considered, using
simulated data corresponding to a variety of instantaneous luminosity
assumptions.  The physics performance differences between the simple algorithm
chosen and the more complicated algorithms, all of which removed the origin
constraint, were negligible.  The more complex algorithms resulted in
processing times two and three times larger than the algorithm above. 

\subsection{Trajectory Fitting\label{fit}}
After the pattern recognition is completed, the SMT clusters selected during
the final filtering and the positions of the CTT track at the innermost and
outermost CFT layers are used as inputs to a $\chi^2$ fit to a circular
trajectory.  

\begin{figure}
\insertyfig{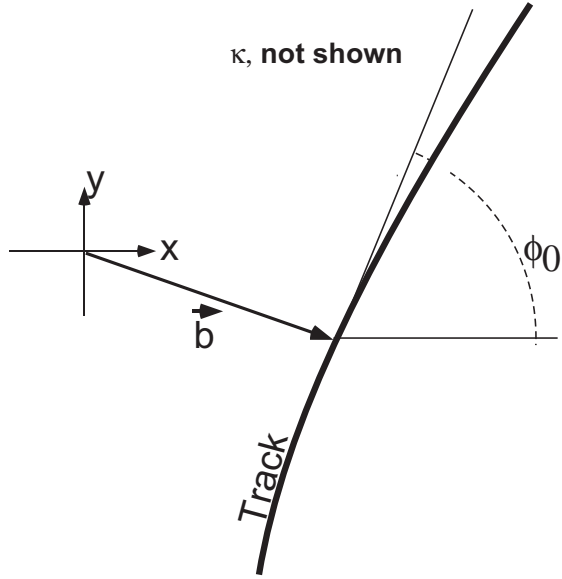}{3truein}
\caption{Illustration of the impact parameter $b$ and track angle $\phi_0$
 used to define the trajectory. The impact parameter sign convention is
 described in the text.  The vector $\vec{b}$ is perpendicular to the
 track where they meet.\label{f-IP}}
\end{figure}
Fig.~\ref{f-IP} shows a single, circular trajectory in
the $r\phi$ plane near the origin.  Three parameters are needed to define a
trajectory, and we use a common definition: (1) the signed impact parameter, $b$,
(2) the angle of the line tangent to the track in the $r\phi$ plane at the
point of closest approach to the origin, $\phi_0$, and (3) the curvature
$\kappa$, related to the radius of curvature $R$ via $\kappa = q\times
B/(2R)$ and inversely proportional to the track momentum in the $r\phi$ plane.
Here $q$ is the particle charge and $B$ is the $z$--component of the D\O\
magnetic field measured in Tesla.

The first order approximation (assuming $br<<1$, $\kappa r<<1$ and
$(\phi-\phi_0)<<1$~rad) for a circle segment is given by
\begin{equation}
   \phi(r) = \frac{b}{r} + \kappa r + \phi_0,  \label{e-traj}
\end{equation}
The sign of the impact parameter is based on the position of the coordinate
origin relative to the circular trajectory and on the curvature.  If the
product $b\kappa > 0$, then the coordinate origin lies inside the circle, and
if $b\kappa<0$, the origin lies outside the circle.

The $\chi^2$ function used in the minimization is thus
\begin{equation}
  \chi^2 = \sum_{clusters} \left [ \frac{ r_i\phi_i - r_i\phi(r_i)}{\sigma_i^2}
    \right ]^2 \label{e-chi2}
\end{equation}
in which $(r_i,\phi_i)$ is the position of either an SMT cluster selected in
the final filtering or a cluster on the CTT track and $\sigma_i =
r\times\sigma_\phi$ is the uncertainty in the SMT position measurement in the
$(r,\phi)$ plane.  SMT clusters are required in at least three of the four SMT
layers.  Allowing tracks with clusters from only three layers results in a
significant 
increase in acceptance.  The additional tracks correspond to particles near the
edges of the geometrical acceptance of the SMT or those which cross
non-functioning SMT detector elements. If there are initially four SMT layers
with clusters selected by the final filtering, and if the resulting fit
$\chi^2$ is larger than is acceptable, the cluster with the largest
contribution to the $\chi^2$ is removed from the fit, and the minimization is
repeated.  These are referred to as ``two pass fits'' in the following. The
$\chi^2$ threshold was determined using single muon events of varying
transverse momenta.  The $\chi^2$ function does not account for multiple
scattering, so the threshold is $p_T$ dependent.

Because the $\chi^2$ function is quadratic in the three fit parameters, $b$,
$\kappa$ and $\phi_0$, an analytic minimization can be performed.  The
parameters can then be determined using simple linear algebra.  The parameters
determined by the fit, $b,\ \phi_0,\ \mathrm{and}\
\kappa$, are the primary outputs of the STT.  The hardware implementing
the final filtering and the $\chi^2$ fitting is described in
section~\ref{ss-tfc}.  Performance is shown in section~\ref{s-performance}.

\subsection{Output Data\label{ss-output}}
For each road with sufficient clusters for fitting, the following information is
transmitted to the global L2 processor and is thus available to be used in
trigger decisions:
\begin{itemize}
  \item The impact parameter significance, $b/\sigma_b$,
  \item $b$,
  \item $\phi_0$,
  \item $p_T$, derived from $1/\kappa$ (assuming the $q = \pm e$),
  \item the fit $\chi^2/$degrees-of-freedom,
  \item the number of SMT layers used in the fit, and
  \item a dE/dx value derived from the pulse height values of the SMT hits
    used in the track fit.
\end{itemize}
Miscellaneous additional information is also provided.  The fit data for each
event is formatted into a standard D\O\ L2 record with a three word header
followed by the STT track data, followed in turn by a single word trailer.  In
addition to the STT fit data, the input CTT data are also transmitted to the L2
CTT preprocessor in unmodified form.

Data is also transmitted to L3 for collisions selected by the L2 global
processor.  The STT sends a variety of data.  It includes the STT L2 output
data, the input CTT data and STC clusters.   Occasionally, diagnostic data
are also included in the L3 output as described in the following section.

\subsection{Monitoring and Diagnostic Information}
In addition to the primary functionality, the STT provides monitoring
information required by all components of the D\O\ detector.  The
monitoring data read out request can be made either under the control of an
external CPU or by the D\O\ trigger serial command link.  The latter requests
are synchronized with the event data, and the resulting monitor data must also
be synchronized with event boundaries.

A variety of monitoring data is provided by each STT component.  In addition to
copies of the standard output data, the monitor data includes event counts, CTT
track counts, STT fit counts, processing time information, and data error
counts.  It also includes all found clusters, all clusters associated with
roads in the form resulting from the coordinate transformation described above
and all fit results, including data for roads with less than three layers with
associated clusters.  Hardware state histograms and processing times are also
available.

\section{Silicon Track Trigger: Hardware Overview\label{s-stt-hardware}} 


The STT hardware consists of custom and commercial electronics mounted in
standard 9U$\times$400~mm VIPA/VME64 crates~\cite{VIPA}.  Symmetry in the D\O\
detector construction and read out implies that six identically populated
crates (sextants) can be used, each receiving complete data from a 60$^o$
$\phi$ wedge of the SMT detector.  No STT data is communicated between crates;
each sextant functions independently and in parallel.  The STT processing
within a sextant is subdivided in to two independent sectors, of 30$^o$ each,
making a total of 12 independent STT sectors. A 2\% acceptance loss arises from
tracks that straddle sector boundaries and thus cannot be reconstructed.

\begin{figure}
\insertfig{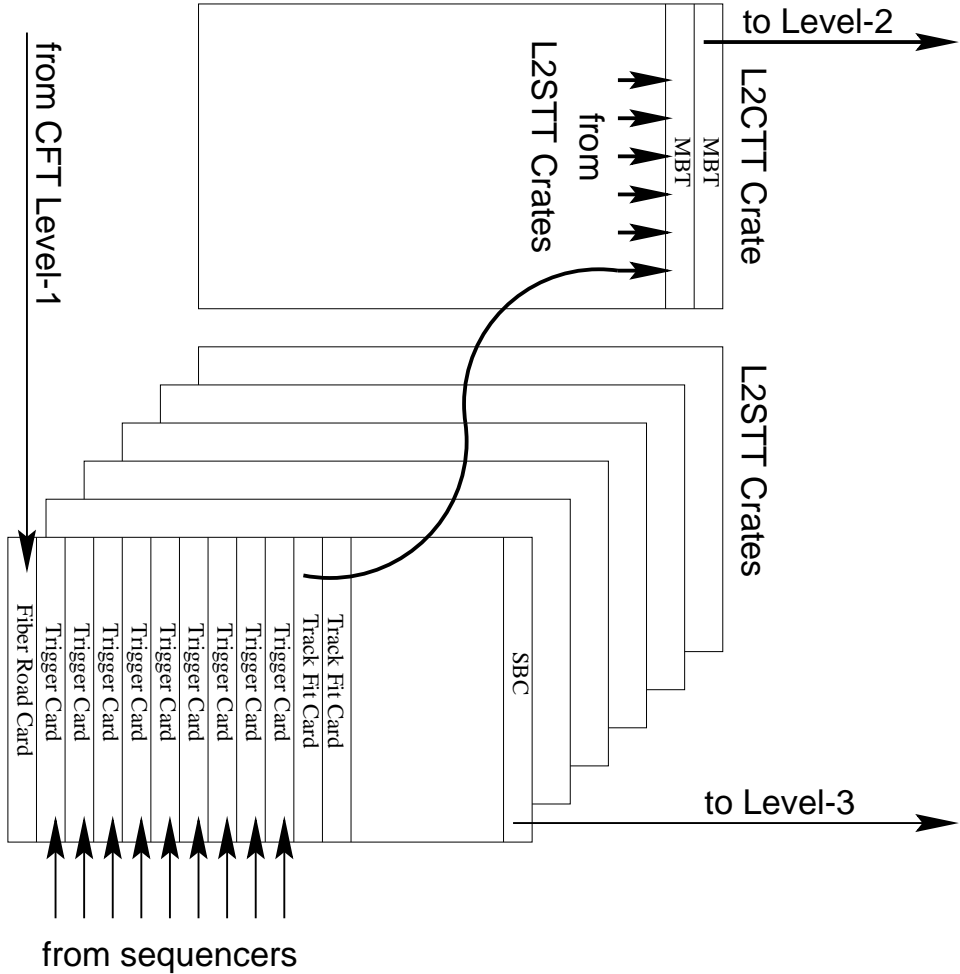}
\caption{The layout of an STT sextant crate. \label{f-crate}}
\end{figure}
A sextant crate is shown in Fig.~\ref{f-crate}. A crate consists of two
different, slightly enhanced commercial CPU boards, and 12 custom processor
elements.  The custom boards are implemented as daughter boards mounted on a
common motherboard design.  In addition, the rear side of the backplane is used
to hold VME transition modules (VTM)~\cite{VTM}, optical receivers used by D\O\
for the CTT and SMT data. The main electronics modules in a sextant are:
\begin{itemize}
\item  One Fiber Road Card (FRC).  This board receives
 experiment-wide synchronization signals via an SCL, maintains internal
 synchronization, and receives and distributes the CTT track data.  The
 FRC distributes these data in parallel to both STT sectors in the crate. The
 FRC hardware is described in detail in section~\ref{ss-frc}.
\item Nine Silicon Trigger Cards (STC).   The STCs receive raw,
 zero-suppressed data from the SMT detector and the CTT tracks from the FRC.
 The STCs perform the silicon cluster finding and initial pattern recognition
 described in section~\ref{s-stt}.  The STC hardware is described in detail in 
 section~\ref{ss-stc}.
\item Two Track Fit Cards (TFC).  The track fit cards receive CTT track data 
 from the FRC and cluster data from the STCs.  The TFC performs the final 
 filtering and track fitting described in section~\ref{s-stt}.  It also sends 
 the fit results to the L2CTT processor.  The TFC hardware is described in
 detail in section~\ref{ss-tfc}, and performance figures are given in 
 section~\ref{s-performance}.
\item A Crate Control CPU (not shown).  VME crate control is provided
 by a Motorola MVME2302 CPU.  It is also used for STT configuration and 
 control and for monitoring-data transfers. 
\item L3 read out is provided via a second commodity CPU board.  This CPU, 
 called the Single Board Computer (SBC),  communicates with the buffer 
 controllers (BC) present on each motherboard
 (see section~\ref{ss-mb}) and is part of the standard D\O\ read out chain.
\end{itemize}

Data communication between FRC, STC and TFC is primarily over low--voltage
differential serial (LVDS) links~\cite{LVDS}, which transmit 32~bit words at
32~MHz.  Output is sent to the L2 CTT preprocessor from the TFC using the
Cypress Hotlink protocol~\cite{b-cyhot} at 160 Mbs, with D\O\ defined data
formats providing event boundary and data validity information.  L3 data is
read out only for events accepted by the global L2 trigger. This slower, less
frequent read out is performed over the sextant backplane VME bus by the SBC.

As previously mentioned, the STT custom modules use a motherboard/daughterboard
design.  The motherboard has sites for: (1) a single main logic daughter card
(FRC, STC, or TFC), (2) up to six I/O cards, either LVDS link transmitter
boards (LTB), LVDS link receiver boards (LRB) and/or Hotlink transmitter
boards, and (3) one L3 read out buffer controller board.  The motherboard
connects these daughterboard sites with three peripheral component interface
(PCI) buses~\cite{PCI} that are connected by a bridge to the VME bus on the
backplane. The motherboard also provides connections to the CTT and SMT input
data via dedicated pins through the backplane.  In addition to providing
modularity by separating the logic functions from the I/O functions, designing
cards that require only standard PCI bus interfaces allowed the purchase of
commercial PCI programmable logic cores~\cite{megacore} and permitted
considerable debugging independent of the VME system and motherboard.  The
motherboard is described in detail in section~\ref{ss-mb}, the LVDS
transmitters and receivers in section~\ref{ss-lxb}, the Hotlink transmitter in
section~\ref{ss-hotlink} and the L3 buffer boards in section~\ref{ss-l3}.

\subsection{Data Flow}
\begin{figure}
\insertyfig{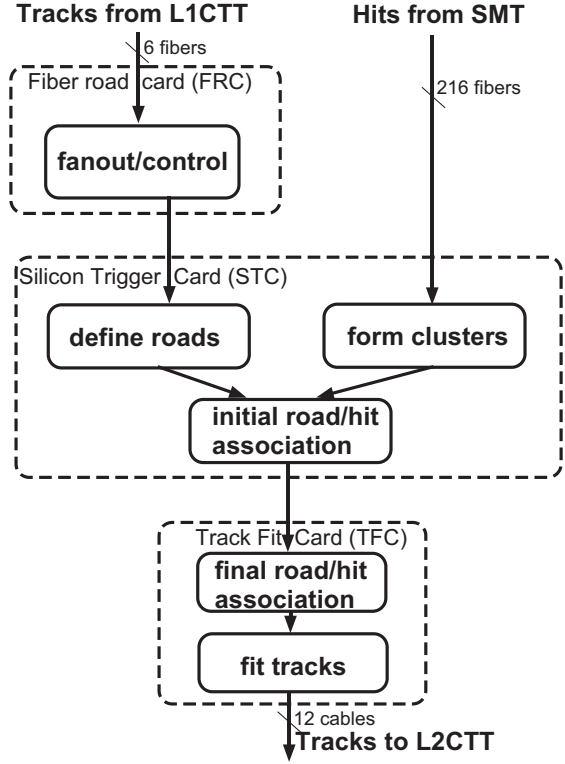}{4truein}
\caption{Data flow through the STT system.  Data paths to the L3
  trigger are not shown. \label{f-sttflow}}
\end{figure}

The data flow through the STT, illustrated in Fig.~\ref{f-sttflow}, is designed
such that only the FRC requires direct event synchronization with the D\O\
trigger system.  

The receipt of a L1 trigger accept is indicated to the FRC by data received on
the serial command link.  The FRC then looks for data from the CTT.  The
Tevatron turn and bunch crossing numbers\footnote{The turn and bunch crossing
numbers allow a given beam crossing to be assigned a number which is unique
over a large enough time interval to provide synchronization checks.} in the
CTT data stream are compared with those SCL data to ensure event synchronization.
The CTT data is then sent by the FRC in parallel to the STCs and TFCs in the
same crate over LVDS links.  For all LVDS channels, transmission of event data
starts with a dedicated header, then the event data, and the transmission ends
with trailer data. A channel without data for a given event must still send
header and trailer words to maintain event synchronization.

The CTT and SMT input data are transmitted to the STT using the G-Link
fiber-optic protocol.~\cite{b-glink} 
The data are received by VTMs in the back of the crate, and then passed
through the backplane to the motherboard (and on to the logic daughter cards)
on dedicated user-defined pins.

STC event processing is initiated when CTT data are received from the FRC or
when SMT data arrives from the detector read out. An STC waits for SMT
data to appear on its inputs within a certain time window around the receipt of
the FRC(CTT) data. Each STC processes eight independent channels of SMT data
in parallel, reading the data, performing on-the-fly clustering and CTT road
association, and then sends the clusters associated with each road to a TFC in
the same crate over a single LVDS output channel.

The TFC begins reading the FRC and STC data as soon as the FRC data for one
event is complete.  After the FRC data are read, the TFC begins reading the
data from each of the STCs.  If STC data on a given channel is not available
immediately, the TFC waits for the complete event data to appear before moving
to the next channel.  When the TFC is finished with all fits from a given
event, the results are transmitted to the L2CTT processor via a Hotlink
transmitter.

The L2CTT processor is a standard D\O\ L2 system module~\cite{b-d0b}.  It
receives the CTT input tracks and fit data from all STT crates, reformats the
data, sorts it into two lists, and then transmits the results to L2 global.
The first of the two lists contains STT and CTT tracks, and it is sorted by
decreasing track $p_T$.  All CTT tracks which gave rise to successful STT fits
are discarded from this list.  The second list contains only STT tracks, and it
is sorted by decreasing impact parameter significance.

Data for L3 read out are transferred from all logic boards to buffer 
controllers for every event accepted by the L1 trigger.  If a given 
event is selected by the L2 trigger, the corresponding
data in the buffer controller are subsequently read out to L3.  If the
event is not selected by the L2 trigger, the buffer 
for that event is released.


\section{Silicon Track Trigger: Hardware Details\label{s-stt-detail}}

\subsection{Motherboard\label{ss-mb}} 
%

\begin{figure}
  \insertfig{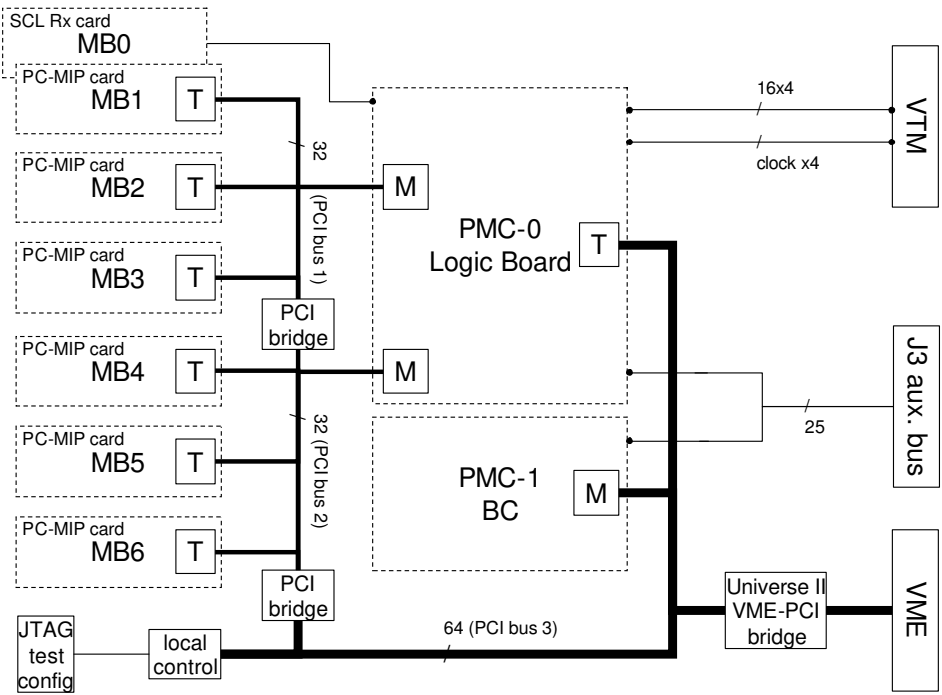}
  \caption{Block diagram of the motherboard. \label{fig:mb1}}
\end{figure}

\begin{figure}
  \insertfig{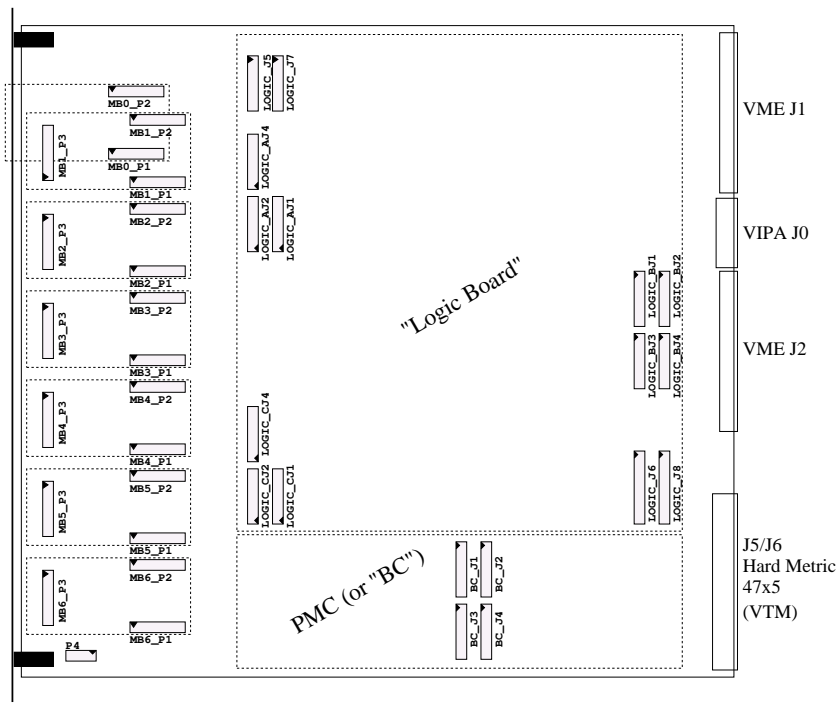}
  \caption{Connector locations on the motherboard. \label{fig:mb2}}
\end{figure}

The motherboard forms the basis of all custom-built STT electronics modules. It
is designed to comply wherever possible with the VME64, VME64x and VME64xP
(VIPA)~\cite{VIPA} standards. The motherboard is a 9U$\times$400mm multilayer
circuit board as defined mechanically in ANSI/VITA 1.3~\cite{VITA}. It has P0,
P1, and P2 connectors, with pinout as specified in VME64xP. The J5/J6 connector
combination is the 2~mm hard metric 47$\times$5 format connector suggested in
VME64x. It is designed to mate with the SVX-type J3 backplane as implemented at
Fermilab, for communication with the transition modules in the rear card
cage. Fig.~\ref{fig:mb1} shows a block diagram of the motherboard and
Fig.~\ref{fig:mb2} shows the connector locations.

The motherboard has sites for up to nine daughterboards. MB0 can hold a D\O\
SCL receiver~\cite{SCL}, while MB1--MB6 are standard PC-MIP~\cite{PCMIP}
sites. MB0 and MB1 cannot be occupied simultaneously. The ``Logic Board'' site
holds a large multi-PMC type board. The ``BC'' site is a standard
PMC~\cite{PMC} site.

The PC-MIP sites are designed to comply with the PC-MIP draft
specification~\cite{PCMIP}. These sites are designed to be used for
point-to-point link drivers and receivers. They are Type II PC-MIP sites (front
panel I/O) with the optional J3 connector provided. Briefly, the PC-MIP
standard defines a pinout for a 3.3V, 32 bit 33 MHz implementation of the PCI
bus~\cite{PCI} on the J1, J2 connectors, and permits 50 user I/O signals on
the J3 connector. The J3 user I/O signals are used for non-PCI communication
between mezzanine cards. The JTAG~\cite{JTAG} interface as defined on PC-MIP
is driven by the motherboard local control ASIC and may be used to provide JTAG
services to the PC-MIP sites.

The ``Logic Board'' site uses multiple PMC sites~\cite{PMC} with extra
connectors added. The connectors labeled LOGIC\_ xJx are in the standard
arrangement for the PCI bus on a PMC card. Thus, in principle, one could mount
up to four standard PMC cards on a motherboard.

The connectors P1 and P2 of the MB0 site that accommodates the SCL receiver are
mapped one-to-one to connectors J1 and J5 of the ``Logic Board''
site. Connections to a dedicated bus implemented on VME J3 row C, as per
D\O-standard (see Ref.~\cite{VRB}) is made through connectors LOGIC\_BJ4 and
BC\_J4. Connectors J6 and J8 of the ``Logic Board'' are routed to VME J3
transition module connections~\cite{VTM}. Some local buffering of signals is
also provided on the motherboard.

Three separate PCI buses are implemented on the motherboard. All buses are
specified to operate at 33~MHz. Buses 1 and 2 connect the ``Logic Board'' site
to the PC-MIP sites. MB1--MB6 are fixed at 32 bits wide due to PC-MIP
limitations. Bus 3 is 64 bits wide on the motherboard and connects the ``Logic
Board'' and ``BC'' sites to a Tundra Universe-II chip~\cite{U2} that provides a
complete VME-to-PCI bridge. The VME side provides all VME64 modes except A64,
allowing 64-bit transfers between the VME and PCI buses. All PCI buses on the
motherboard normally use 3.3V signaling. However, because the Universe-II chip
only supports 5V signaling, the devices connected to PCI bus 3 must be
5V-tolerant.

A PCI target interface is provided on the motherboard (connected to PCI Bus 3)
for local control and monitoring. This interface provides access to FPGA
programming resources, JTAG boundary scan access to mezzanine boards, and
access to the VTM serial control bus via J3.

\subsection{Fiber Road Card\label{ss-frc}} 
%

The Fiber Road Card (FRC) serves as the main communication link
between the STT and the rest of \Dzero .
It has five main functions:
\begin{enumerate}
  \item Communicate with the \Dzero\ trigger timing and control system
	(the Serial Command Link Hub)
	and initiate any action requested by it.
  \item Receive tracks found by the CTT.
  \item Distribute trigger information and CTT tracks to the STCs
	and TFCs.
  \item Control the transfer of data produced in the STT system to the
	L3 system on L2 accepted events.
  \item Perform special VME bus request arbitration to resolve
	conflicts between the in-crate CPU, used for initialization
	and monitoring data collection, and the Single Board Computer
	(SBC), used to collect L3 data.
\end{enumerate}

\begin{figure}
  \insertfig{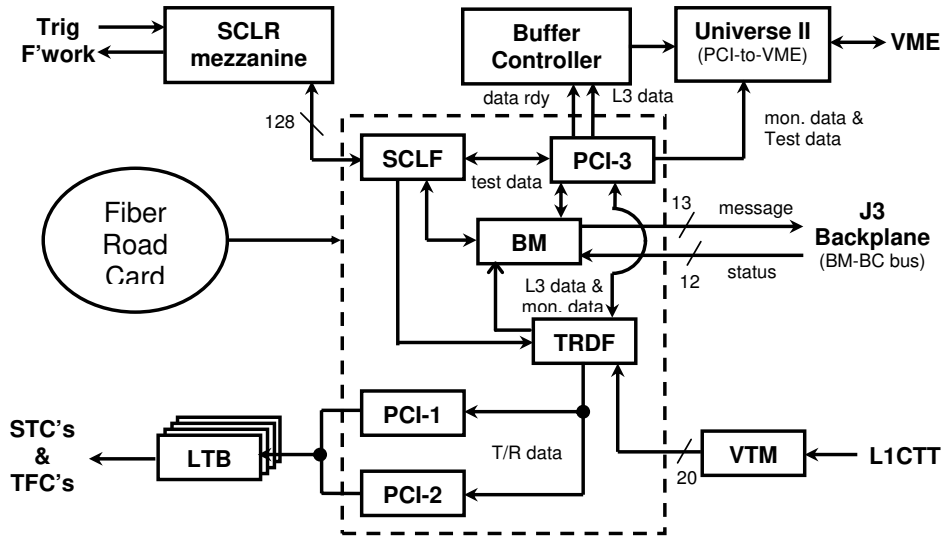}
  \caption{A block diagram of the main functional elements of the
	   FRC (enclosed in the dashed rectangle). \label{fig:frcblk}}
\end{figure}

The FRC is implemented as a PMC daughter board on the standard STT
motherboard described in section~\ref{ss-mb}.
A block diagram of the FRC daughter board functionality is given in
Fig.~\ref{fig:frcblk}.
It is broadly composed of four functional elements
that are implemented in six Altera FLEX 10K50 or 10K30 FPGAs
\cite{altera}:
\begin{enumerate}
  \item The SCL Formatter (SCLF), providing communications with the
        SCL via an SCL Receiver mezzanine card.
  \item The Trigger/Road Data Formatter (TRDF), 
	receiving CTT data from a VTM 
	and SCL information from the SCLF, 
	and constructing Trigger/Road data blocks 
        that are transmitted to the STCs and TFCs.
  \item The Buffer Manager (BM), controlling the
        allocation/deallocation of all L3 data buffers in the system and
        controlling requests for data transmission over the VME bus.
  \item PCI interfaces (PCI-1,2,3), controlling the transfer of all
        data off of the FRC daughter board.
\end{enumerate}
Each of these elements is described in more detail in the following sections.

\subsubsection{Serial Command Link Formatter\label{ss-sclf}}
Control and timing
signals from the \Dzero\ trigger framework \cite{framework}
come to the FRC from
the Serial Command Link (SCL) hub on coaxial 1-Gbs cable
(Times Microwave LMR-200)
and status information is returned to the hub on RS485 26 conductor cable
\cite{framework}.
This data is received/transmitted by an SCL Receiver (SCLR) mezzanine board
\cite{SCL} that plugs into a special PCI MIP site (MB0) on the
motherboard. 
Information from the SCL hub is presented to the FRC every 132 ns
as 128 bits on two 64-pin connectors.
Status information sent back to the hub is implemented as voltage
levels on the RS485 cable.

The SCL Formatter (SCLF), implemented in an Altera Flex 10K50 chip on
the FRC, picks off SCL information relevant to the STT and directs it
to the appropriate elements on the FRC for distribution to the rest of
the system. Distributed information includes:
the accelerator clock;
L1 accept and L2 accept and reject bits;
the Tevatron bunch crossing (BX) and turn (TURN) numbers associated with the
current L1 or L2 information;
trigger qualifiers indicating special actions associated with the
current event, such as a request to switch to ``full read out mode'' 
or a request to collect online monitoring data;
the system initialization request from the SCL
and various busy and error signals.

\subsubsection{Trigger/Road Data Formatter\label{ss-trdf}}
Data from the CTT is received at the FRC by the VRB Transition
Module (VTM) \cite{VTM} on a single G-Link optical cable, serially at
1.06 Gbits/s. 
Each word from the CTT consists of 16 data bits and four bits used
as control characters to flag the first and last words in an event.
The VTM parallelizes this serial data and presents it,
at 53 MHz,
on 20 data pins
at one of the four VTM input channels on the
motherboard J3 connector.

The Trigger/Road Data Formatter, implemented in an Altera Flex 10K50
chip on the FRC, receives this CTT data as well as SCL information
from the SCLF. It checks for a BX number mismatch between the
two data sources, informing the SCLF to request an initialization from
the SCL hub if a mismatch is found. If the data are consistent, the
TRDF then constructs the Trigger/Road (T/R) data block, which consists
of headers and trailers containing trigger and status information,
surrounding the unmodified CTT data. The T/R data block, which
contains between six and 54 32-bit words, is sent to PCI-1 and PCI-2
on the FRC for transmission to the STCs and TFCs via LTBs.

\subsubsection{Buffer Manager\label{ss-bm}}
Distribution of control signals within the STT system is accomplished
by two means: (a) ``trigger qualifier'' bits included in the T/R data
header by the TRDF and sent to the STCs and TFCs, 
and (b) logic in the Buffer Manager (BM) that coordinates
communication with all external requesters of STT data (aside from
L2CTT, which receives its data from the TFC).
This scheme centralizes distribution of 
SCL status information and
requests for monitoring and L3
data and for initialization within the STT system,
allowing the design of only one interface to these external elements.
It requires that the BM communicate with 
the SCL hub via signals passed to/from the SCLF,
the in-crate CPU via several VME interrupts,
the SBC via user-defined lines on the VME backplane
and the Buffer Controllers (see section~\ref{ss-l3}) on each of
the boards in the system via a custom bus on the J3 backplane.

The BM has six main tasks.
\begin{enumerate}
  \item The BM generates ``error'' and ``busy'' signals based on
	information from all elements of the system. This information
	is passed to the SCL hub (through the SCLF) and allows the
	STT system to request re-initialization if it detects a fatal
	error or to request that L1 accepts be halted if it falls
	behind in processing its data.
  \item It passes initialization commands from the SCL (SCL Init) to
	the in-crate CPU as a VME interrupt and informs the SCL hub of
	the status of its request. 
	The CPU notifies all other boards in the system of the
	SCL Init request by setting VME-accessible registers.
  \item It informs the CPU that it has received a ``collect monitoring
	data'' request from the SCL by sending an interrupt. The CPU
	then manages the collection of monitoring data from
	VME-accessible memory on each board.
  \item It manages the buffering of L3 data in each board's Buffer
	Controller (see section~\ref{ss-l3}).
  \item It informs the SBC when all L3 data in the system is ready to
	be read out following the receipt of an L2 accept.
  \item It schedules VME accesses in the system during normal data
	taking by controlling when SCL Init, Monitoring requests and
	L3 data transfers are initiated.
\end{enumerate}

The logic for all of these tasks is contained in an Altera Flex 10K50
chip. 

\subsubsection{PCI Interfaces\label{ss-frcpci}}
All data transfer to and from the FRC daughter board,
with the exception of SCL data and CTT roads,
is accomplished using the three PCI buses (PCI-1,2,3).
Interfaces to each of these buses on the FRC side are
implemented in Altera Flex 10K30 (PCI-1,2) and 10K50 (PCI-3) chips
using Altera ``Megacore'' firmware \cite{megacore}.
The PCI-1,2 buses are used for the transfer of identical copies of the
T/R data block to four LTBs on the FRC motherboard.
They function as PCI masters when transmitting this data and can also
function as PCI targets for test data read out.
The PCI-3 bus is used for communication with the Buffer Controller and
for most VME interactions.
The FPGA containing its interface has only PCI target functionality,
but also
includes buffers for L3 data and for data collected by the online
monitoring system.

\subsection{Silicon Trigger Card\label{ss-stc}} 
%

\subsubsection{General Overview}
The Silicon Trigger Card (STC) is the second logic board flavor. It receives
the raw SMT data, performs pedestal subtraction, non-functioning and noisy
channel masking, forms clusters and associates these clusters with roads
defined by the seed tracks from the CTT.

The STC board is a multilayer circuit board that 
plugs into all the ``Logic Board'' site connectors on the motherboard except J5
and J7.

\begin{figure}
  \insertfig{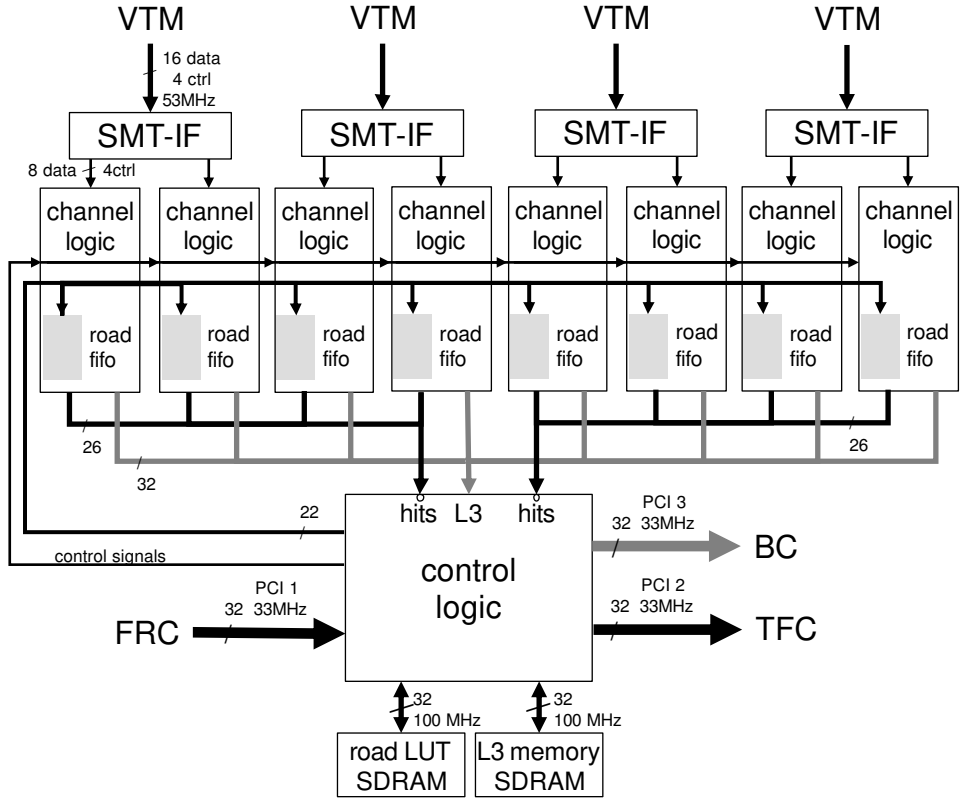}
  \caption{Block diagram of the STC. \label{fig:stc}}
\end{figure}

The STC receives the following inputs:
\begin{itemize}
\item CTT seed tracks from the FRC via PCI-1.
\item raw data from the SMT via optical fibers using the HP G-link protocol. 
 Four fibers plug into optical receivers on a VTM\cite{VTM} in the rear card cage. Each
 fiber carries the data from two SMT detector elements. Thus each STC processes
 the data from eight SMT detector elements. The data are converted into
 electrical signals and transmitted through the J3 backplane to connectors J6
 and J8 on the STC board. The data from each detector are eight bits wide. In
 addition there are four control bits.
\item downloadable parameter tables and firmware from the CPU board via the 
  VME bus and PCI-3.
\end{itemize}
The STC has the following outputs:
\begin{itemize}
\item clusters associated with the seed tracks via PCI-2.
\item diagnostic data for read out by the data acquisition system via PCI-3.
\end{itemize}
The main logic is programmed into a single large Xilinx Virtex
 FPGA\cite{b-xilinx}. It consists of the following elements: SMT input FIFOs,
 channel logic, and control logic. Fig.~\ref{fig:stc} shows a block diagram
 of the STC logic.

\subsubsection{SMT Input FIFO}
The four SMT input FIFOs (SMT-IF) each receive the data from one optical fiber
from the VTM. The data are 20 bits wide, eight bits from each detector element
and four control bits. They are strobed into the SMT-IF by a 53 MHz clock from
the VTM. The SMT-IF is an asynchronous FIFO and splits the data into two
streams consisting of eight data bits and four control bits to be fed into the
channel logic. In test mode, the SMT-IF blocks also can hold simulated SMT data
to test the STC logic in a stand-alone way.

\subsubsection{Channel Logic}
The eight channel logic blocks each process the data from one SMT detector
element. They perform pedestal and gains corrections, combine data from
adjacent strips into clusters, calculate the cluster centroids, and associate
the clusters with the seed tracks. These hits are accumulated in a FIFO for
later read out by the control logic. Another FIFO accumulates diagnostic data
for later read out. Each channel logic block consists of three sub-blocks:
strip reader, centroid finder, and hit filter.

The strip reader receives the data from the SMT-IF. It usually sits in its idle
state waiting for data. The silicon data are digitized on the detector by SVX
II chips~\cite{SVXII} which contain 128 channels of analog pipelines and 8-bit
ADCs. Each channel is connected to a strip on the silicon detector. 
The data read out from the detector has been sparsified. The data consist of
SVX chip id followed by two eight-bit numbers indicating strip number and data
value for each strip above threshold. As the data are received chip-by-chip
pedestal and gain corrections are performed. Data from strips that are marked
bad are set to zero. The pedestals, gains, and bad strip lists are stored in
lookup tables that are downloaded at initialization time via PCI-3. When the
strip reader encounters the end-of-event record, it goes back to its idle
state.

The SMT data does not contain synchronization information to identify the beam
crossing it originates from. Thus synchronization with the seed tracks from the
CTT is achieved by a time coincidence between the arrival of the CTT data and
the SMT data at the channel logic. The CTT data arrives from the FRC via PCI-1
and is sent by the control logic to all channel logic blocks. Arrival of either
CTT or SMT data starts a counter that counts down from a downloadable starting
value. The maximum delay that can be programmed is 8$\mu$s. If the other data
are received before the counter reaches zero, a trailer is inserted that
contains the event number from the CTT data. If the SMT data are missing, an
empty event record is generated and if the CTT data are missing the SMT data
are merged with the following event.

The pedestal-- and gain--corrected SMT data are then fed into the centroid
finder at a rate of 25~MHz. The centroid finder runs at 50~MHz so that it always
keeps up with the data arriving at its input. The cluster finder scans through
the strips forming clusters according to the algorithm defined in section
\ref{ss-clustering}. Numerator and denominator for the centroid calculation are
accumulated as the strips are processed and the quotient is calculated as soon
as the cluster is complete. The denominator, which is the total pulse height of
the strips used in the centroid calculation, is encoded in three bits using a
table of thresholds. The centroid position is encoded in 13 bits. Four bits
identify the SVX chip, seven bits identify the strip, and two bits specify the
position to quarter--strip precision. The centroid position and the three-bit
pulse height are stored in a FIFO that can accommodate up to six events.
Because the pulse height is a measure of the energy deposited by the particle
traversing the SMT, it is transferred to the TFC and then to the L2CTT.  In
order to save data transmission the dynamic range is limited to three bits of
encoded pulse height information.\footnote{The pulse heights from all clusters
used when fitting a track (Section 4.4) are a measure of the energy lost in the
SMT by the particle giving rise the track, and thus could provide particle
identification.  This capability is under study.}

The hit filter compares the centroid positions with the address ranges that
define the roads. For each seed track, the road FIFO is loaded with two 11-bit
addresses defining chip and strip numbers of the upper and lower edges of the
road defined by this seed track. The STC was designed to handle up to 48 roads
per event. Typically the number of roads is much smaller, one or two per STT
sextant. Every centroid is sequentially compared to every road. If the leading
11 bits of the centroid position lie between the two road edges, the centroid
is associated with that road. We call centroids that are associated with a road
``hits''. All hits, consisting of the 16 bits of centroid data and the six-bit
road number, are stored in a FIFO. A centroid can be associated with more than
one road and therefore give rise to more than one hit. Filtering starts when
all roads have been loaded into the road FIFO and at least one centroid is
present and proceeds at the speed of the system clock (100MHz). Thus every 10
ns a centroid is compared to a road. Hits produced by the channel logic blocks
are transferred to the control logic over two 26-bit buses.

For diagnostic and monitoring purposes, the channel logic can accumulate
various data that it receives or generates in FIFOs to be read out by the
control logic. Which data are stored is determined by a downloadable
configuration word. Possible data types are uncorrected and corrected SMT data,
axial clusters and centroids, z--centroids, event error flags and data marked
as bad by the noisy/bad channel flags. We refer to these data as ``L3 data''
because they are intended to be transferred to L3 upon L2 accept of an event.
Not all the data from every detector in a given STC is written on each event.
Instead, data from one SMT detector per STC is written for each event with the
chosen detector changing each event.

\subsubsection{Control Logic}
The control logic block manages the processing of the event by the STC. It
provides the interfaces with the PCI logic and issues control signals to the
channel logic. Its components are the interfaces with PCI-1 and PCI-3, the
L2 logic, and the L3 logic.

The PCI-1 interface receives the data from the FRC, extracts the event number
and converts the seed tracks into address ranges by addressing a lookup table
in an external 64 Mb synchronous dynamic RAM (SDRAM). The range limits are
loaded into the road FIFOs in the channel logic blocks.

The PCI-3 interface receives all the downloadable parameters and transfers the
L3 data to the buffer controller. It also contains the logic that resets the
STC when an initialization request is received from the trigger framework (SCL
Init).

The L2 logic formats the list of hits for transfer to the TFC and sends it out
via PCI-2. The L3 logic collects the L3 data from the channel logic and stores
it temporarily in an external SDRAM until the data are transferred to the
buffer controller via PCI-3 when PCI-3 is available.

\subsection{Track Fit Card\label{ss-tfc}} 
\begin{figure}
\insertfig{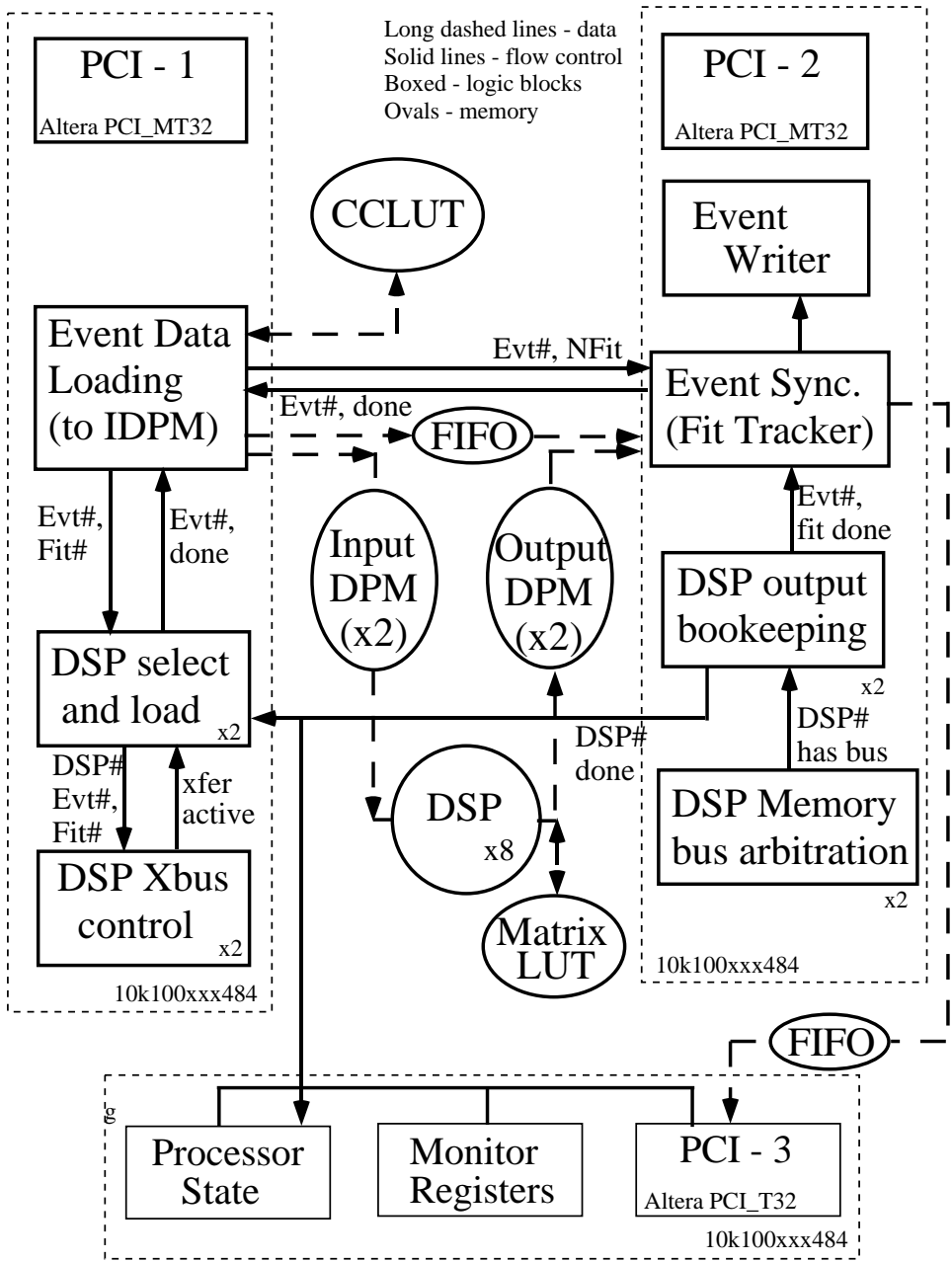}
\caption{Data flow through the TFC.  Dotted lines indicate data paths; solid
lines, control paths.  Boxes represent logic functions and the boxes with
dotted lines indicate specific Altera 10k100 FPGAs.\label{f-tfc1}}
\end{figure}

\begin{figure}
\insertyfig{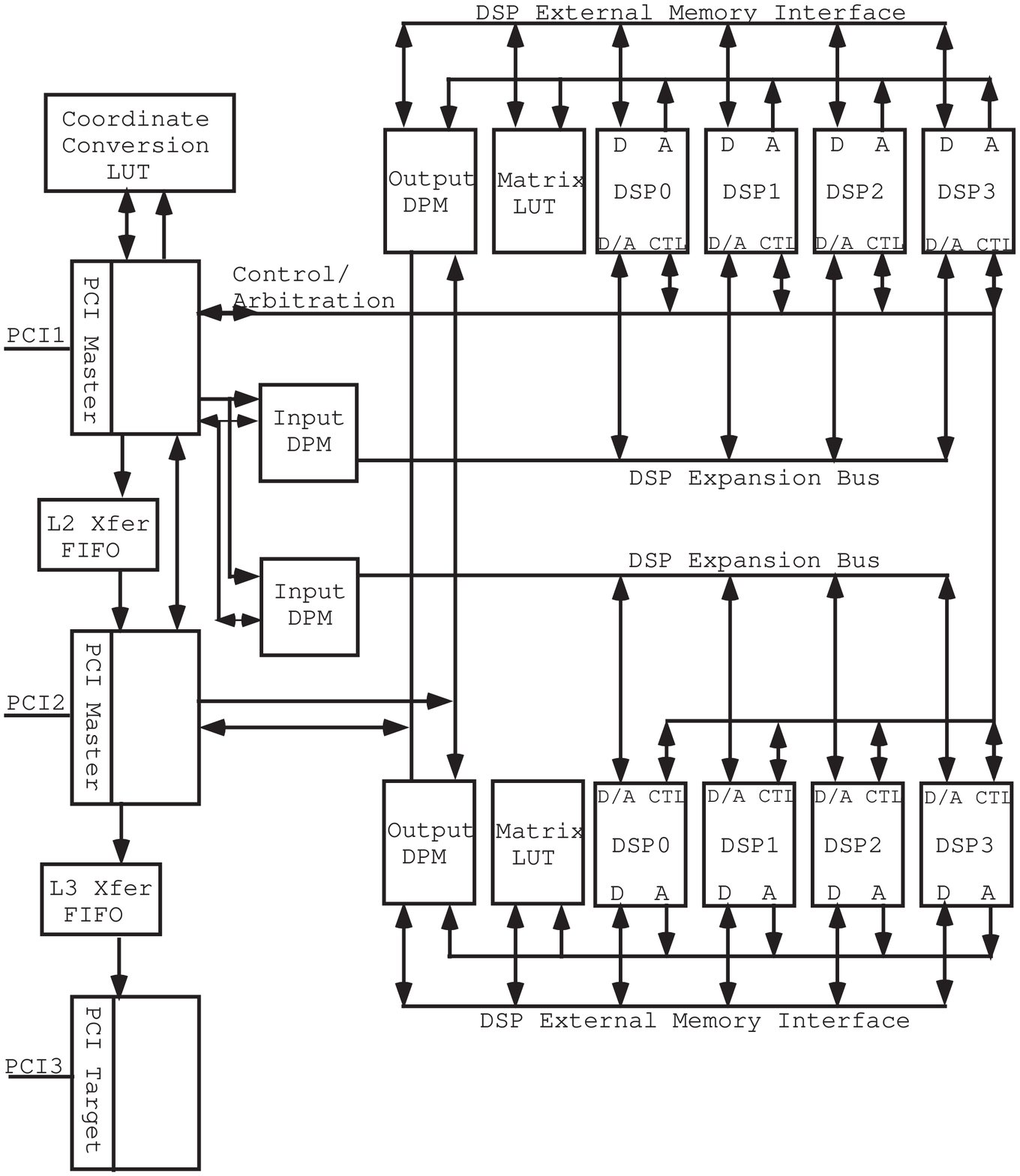}{5truein}
\caption{A block diagram of the major functional elements of the TFC, including
the separate buffer memories for each of the two sets of four
DSPs.\label{f-tfc2}}
\end{figure}

The track fit card (TFC) provides three parts of the STT functionality
described in section~\ref{s-stt}: (a) the final filtering step of the pattern
recognition, (b) the trajectory fits to determine $b$, $\phi_0$ and $\kappa$
and (c) the output to L2.  The overall hardware design uses programmable logic
for data flow control, buffer control and processor scheduling, and digital
signal processors (DSPs) for the numerical calculations needed for final
filtering and track fitting.  The TFC is described in three parts: (1) the
input data sources and buffering hardware, (2) final pattern recognition and
trajectory fitting hardware and algorithm details, and (3) output buffering
hardware.  Overviews of the TFC internal data flow and hardware structure are
shown in Fig.~\ref{f-tfc1} and Fig.~\ref{f-tfc2} respectively.

\subsubsection{TFC Input Data and Buffering}
A track fit card receives CTT track data from the FRC via an LVDS channel.  It
also receives SMT hit data from six STCs via LVDS links, with one channel for
each STC.  Because of the SMT geometry, data from three of the nine STCs are
sent to both TFCs in one crate.  Data are received on all seven LVDS inputs for
every collision selected by the L1 trigger.  The data from the seven channels
are read sequentially on the PCI-1 bus on the motherboard.  The PCI-1 bus is
dedicated to reading input data; no other data routinely passes over this bus
during data taking.

The CTT data are an unaltered copy of the data received by the FRC, with the
internal LVDS header and trailer added.  The STC data for each hit has
the hit centroid, the encoded pulse height and an index identifying which of
the CTT roads the cluster has been associated with during the initial pattern
recognition phase.\footnote{If a cluster is associated to multiple CTT tracks,
then it will be sent once for each track it is associated with.}  The index
number corresponds to the order in which the CTT tracks occur in the input; a
cluster with road index 12 is associated with the 12$^{\mathrm{th}}$ CTT road
in the given STT sextant.

   The SMT to STC cabling implies that clusters from the same trajectory can
be spread across the six STC channels.  Because of this, the processing of a
given event in the TFC must wait until the data from all STC inputs have been
read.  The road and cluster data are initially held in input buffer dual--port
memory (IDPM) on the TFC.  The buffer memory is organized such that the CTT data
and the SMT clusters for a given road are held in consecutive memory locations.
The complete data for each road can be up to 58 words.  A specific region of
buffer memory containing 64 locations is permanently assigned to each possible
road in an event.  The memory has enough room to hold up to 64 roads for each
of 16 events.  The buffer space is divided into two independent memory banks,
one of which stores the data for the odd--numbered roads in an event and the
other of which stores the data for the even--numbered roads.  This split is
made to reduce bus contention when loading the data into the processors.  The
buffers use dual port memory, and data for one event can be written into the
input buffers at the same time as data for a different event is being read out
of the buffers for processing.

   During the initial read of the STC data from the LVDS receivers into the
input buffer memory, an on-the-fly conversion of the cluster position from the
encoded form to the format used in the fitting is performed.  The conversion is
provided by using the encoded position data as an address to a 28-bit wide look
up table.  The table output gives the $\phi_i$ position of the cluster relative
to a sector--dependent offset and indices denoting the barrel, layer and ladder
on which the cluster was found.  Full--precision detector alignment constants
are included when creating this look up table.  The information is sufficient
to reconstruct the $(r_i,\phi_i)$ position needed in the track $\chi^2$
calculation.

\subsubsection{Final Pattern Recognition and Trajectory Fitting}
Once all FRC and STC data for a given collision have been loaded into a TFCs
input buffers, the second phase of the processing begins.  Each TFC has eight
Texas Instruments TI320C6203B digital signal processors (DSPs)~\cite{b-dsp}.
The DSPs are organized as two independent groups of four DSPs with each group
having access to input data from one of the input buffer banks.  The two groups
run independently and can simultaneously process data.

When any one of the four DSPs in a group is not processing data, and complete
raw data for at least one road is available in the corresponding IDPM, the
data are loaded into the DSP via its expansion bus (XBUS) under the mastership
of logic implemented in a programmable gate array.  The XBUS transfers 32 bit
words on a 33~MHz clock.  Each group of four DSPs shares a common XBUS, and
bus arbitration is provided by the firmware in the programmable gate array.

Once the transfer has finished, the DSP program, written in C code,
performs the final cluster selection and trajectory fit described in
section~\ref{s-stt}.  The DSP has hardware support for 16-bit integer
multiplies and 32-bit integer sums, but does not provide hardware support for
integer division or for any floating point calculations.  Because of this, the
cluster selection and fitting algorithms are implemented with 16-bit integer
input data.

The final phase of the pattern recognition has been described in
section~\ref{s-stt}.  The modifications required when running in the DSPs
arise because only integer multiplies and adds can be performed efficiently in
the DSP.  To accomplish this, the coefficients $\kappa_{CTT}$ and
$\phi_{0_{CTT}}$ defining the center of the CTT road given by $\phi_{CTT}(r) =
\kappa_{CTT}\cdot r + \phi_{0_{CTT}}$, are not computed in the DSPs but rather all
necessary values of $\kappa_{CTT}$ and $\phi_{0_{CTT}}$ are computed in advance
and loaded into small look--up tables in the DSP internal memory during
configuration.  The CTT input data for each road provides $\phi_i$ indices at
two fixed radii of approximately 20~cm and approximately 50~cm.  Under the
assumptions of cylindrical symmetry and the particle originating at the origin,
the values of $\kappa_{CTT}$ and $\phi_{0_{CTT}}$ can be tabulated as functions
of the $\phi_i$ index at one of the two radii, and the difference between the
two $\phi_i$ indices.  The resulting table is stored in the internal memory of each
DSP.  The assumption that the particle originated from the nominal D\O\
coordinate origin is too simplistic.  An additional set of corrections to the
road center determined above is needed to account for the true beam position.
Corrected values for $\kappa_{CTT}$, $\phi_{0_{CTT}}$ and $b_{CTT}$ are derived
in terms of the known beam position and the initial estimates of the curvature
and azimuthal angle.  The correction factors are derived externally at the
start of each D\O\ run to allow for time variation of the beam position,
converted to a packed integer format and downloaded into internal memory of
each DSP.  The stored values of $\kappa_{CTT}$, $\phi_{0_{CTT}}$, and $\phi_i$
are scaled to provide maximum precision in a 16-bit integer.

Performing a standard, linear least-squares minimization of Eqn.~\ref{e-chi2}
to determine the three track parameters $b$, $\phi_0$ and $\kappa$ gives the
following three equations:
\begin{eqnarray}
 b & = & \Sigma_j M_{1j}\Phi_j, \\
 \phi_0 & = & \Sigma_j M_{2j}\Phi_j, \\
 \kappa & = & \Sigma_j M_{3j}\Phi_j,
\end{eqnarray}
with $j=1, 2, 3$, $\Phi_n = \Sigma_k r_k^n\phi_k/\sigma_k^2$ with
$k=1,..., N_{points}$ and $M_{ij} = f_{ij}(r_i,\sigma_i) = M_{ji}$.
Here $(r_k,\phi_k)$ is the coordinate of an SMT or CTT point used in the fit
and $\sigma_k$ is the resolution of the point.  The functions $f_{ik}$ are
ratios of sums of powers of $r_i$ and $\sigma_i$.

The $M_{ij}$, although simple to write in sum notation, are far too
algebraically complicated to calculate in the DSPs within the L2 time budget.
In addition, the $M_{ij}$ terms have large differences in magnitude so finding
a common rescaling of numerical results into 16 bit integers is difficult.
Finally, the $\Phi_n$ terms involved sums of products of radii and angles,
giving a large number of operations needed to compute the track parameters.

The above result can be rewritten using straightforward algebra as
\begin{eqnarray}
 b & = & \Sigma_k M'_{1k}\delta\phi_k \\
 \phi_0 & = & \Sigma_k M'_{2k}\delta\phi_k + \phi_1\\
 \kappa & = & \Sigma_k M'_{3k}\delta\phi_k
\end{eqnarray}
with $k=2, 3, ..., N_{points}$, $M'_{ik} = \mathit{M_{i1}}r_k/\sigma_k^2 +
\mathit{M_{i2}}r_k^2/\sigma_k^2 + \mathit{M_{i3}}r_k^3/\sigma_k^2$, and
$(\delta\phi)_k \equiv \phi_k - \phi_1$.  This is a significant improvement in
a number of ways.  First, the column vector of measured coordinate residuals
$(\delta\phi)_k$ involves only the angles $\phi_k$.  Because
$\delta\phi_1\equiv 0$, this term is not needed in the above sums.  The second
improvement is that all terms in a given row of the coefficient matrix are of
the same dimension and thus of the same numerical scale.  This allows rescaling
the (real valued) matrix elements $M'_{ij}$ and angles $\delta\phi$ into 16-bit
signed integers while using only one rescaling for the $\phi_k$'s.

In either of the above forms, the DSPs cannot compute the terms $M_{ij}$ or
$M'_{ij}$ with enough precision quickly enough to meet the L2 time budget.
Instead, the entire azimuthal range is subdivided into 1440 $\phi$
sections.\footnote{The number of subdivisions needed was determined empirically
using simulated data.}  Within each of the 1440 sections, the same coefficients
$M'_{ij}$ can be used while retaining sufficient precision in the calculations.

Because the terms in the coefficient matrix depend on which SMT layers
contribute clusters to a given fit and because of mechanical position variations
within the SMT barrels, more than one matrix is needed for each of the 1440
sections.  For example, a given fit in a given section may have four SMT
clusters selected for fitting, the first cluster from layer one, barrel one,
the second from layer two barrel one, the third from layer three barrel one,
and the fourth from layer four of barrel two.  Another fit in the same $\phi$
section may instead have clusters from only three layers, for example layers two,
three and four all in barrel five.  The coefficient matrices for these two fits
will be significantly different.  To have sufficient precision to compensate
for these effects, 8192 matrices are used for each of the 1440 slices giving
roughly 12,000,000 unique coefficient matrices.

All matrices for each $\phi$ slice are computed {a priori}, and the values
are written into look up tables on the TFCs.  Because a given TFC processes
data for only one $30^o\ \phi$ segment\footnote{The actual angular coverage for
CTT input tracks is somewhat larger to include tracks which are outside the
30$^o$ sector at the outermost CFT layer but which curve into the sector.}, each
TFC requires matrices for the 160 $\phi$ slices contributing data to that TFC.
Each matrix has 15 16--bit elements (for which 16 16--bit locations are
allocated for easy address construction).  Thus, 4 MB of memory is needed for
each TFC to hold the precomputed matrices.  Twice this amount of memory is
provided on the TFC, so each group of four DSPs can have its own copy of
the complete set of matrices.  Within a group of four DSPs the matrices are
accessed using a shared data bus, but the two groups are completely independent
and execute in parallel.

\subsubsection{Output buffering and formatting}
As with the input buffer memory, output fit data are stored in dual--port
buffer memory.   Once all fits for a given event are finished, the data is
written to a Hotlink Transmitter using the PCI-2 bus.  On-the-fly formatting,
including addition of a standard D\O\ header and trailer and a checksum is
performed during the PCI write phase.  The TFC is the bus master for this
transfer.  The event time--ordering is preserved during the output writing.


\subsection{Serial Link Transmitter/Receiver Boards\label{ss-lxb}} 
%

\begin{figure}
  \insertfig{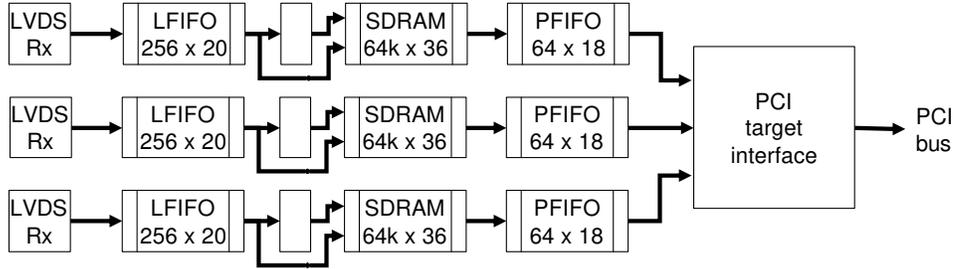}
  \caption{Block diagram of the LRB. \label{fig:lrb}}
\end{figure}

The link receiver board (LRB) has three channels, each capable of receiving
32-bit data words at 33 MHz over 10-conductor category-5 cables that are
connected to 10-pin modular connectors on the front panel. The board has a
slave PCI interface that allows read out of the data over a 32-bit, 33~MHz PCI
bus. Except for the dimensions of the front panel input connectors, the LRB
complies with the PC-MIP standard~\cite{PCMIP} (Type II card). An overall block
diagram of the LRB is shown in Fig.~\ref{fig:lrb}.

The LVDS~\cite{LVDS} receiver is a National Semiconductor DS90CR286 integrated
circuit, a 28-bit receiver which operates at 3{.}3V. The receivers run
continuously at a clock supplied over the link. The clock frequency is 64 MHz,
twice the PCI clock frequency. Each LVDS clock cycle one 24-bit word is
received by the LVDS receiver, consisting of 16 data bits plus control and
error correction bits. Input data is processed through a 5-bit Hamming code
that corrects all single-bit errors and detects all two-bit errors and many
multi-bit errors. The resulting 20 bits (16 data plus control and error flags)
are stored in LFIFO.

LFIFO is a 256-word by 20 bit FIFO. It is written using the LVDS link clock,
and read using the doubled PCI bus clock. The LFIFO output is demultiplexed to
form 36 bit words, which are written to RAM.

The board contains a 256k $\times$ 36 synchronous RAM, which is used to buffer
the data which do not fit in the 256-word LFIFO. The RAM operates at 132 MHz,
four times the PCI clock rate, and is multiplexed to appear as a six-port RAM
(three write ports, three read ports). Typically up to 3 write operations can
be performed in one PCI clock cycle, allowing received data from three active
channels to be stored. A single read operation can also be performed for PCI
access. The SRAM logically appears as three FIFOs, each used by one LRB
channel. The data is read out of the SRAM and written into the corresponding
PFIFO.

PFIFO is a 64-word by 18--bit FIFO. It is written and read using the doubled
PCI clock. It buffers the data between the SDRAM read port and the PCI
interface. Whenever the PFIFO occupancy is not above a certain threshold, data
is read from the SRAM to refill it. Data is read on demand by the PCI
interface.

The PCI target interface block contains the PCI target interface, plus
multiplexing and control logic to merge the data from the three channels into a
single PCI data stream.

The LRB can operate either as three simple receivers with data FIFOs, or in an
event-oriented block mode. In block mode, data is expected to conform to a
specified format with data blocks containing headers with a block ID, optional
data words, and a trailer. Data may be combined across links in such a way that
the next can be read with a single PCI block transfer. Multiple LRBs may be
read out in succession on the same PCI bus with a single PCI block transfer.


Each link transmitter board (LTB) holds a slave PCI interface, formatting
logic, and three LVDS link transmitters. The LTB complies with the PC-MIP
standard (Type II card) to the same extent as the LRB. The three LTB
transmitters operate strictly in parallel and transmit identical data, though
they can be individually disabled.

\begin{figure}
  \insertfig{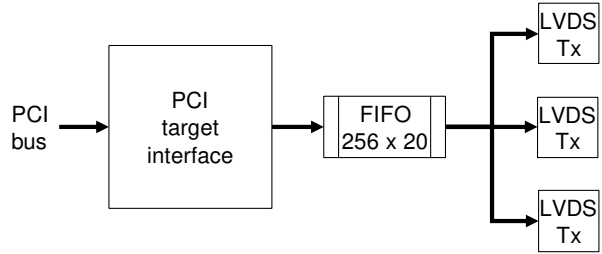}
  \caption{Block diagram of the LTB. \label{fig:ltb}}
\end{figure}

Data is written as 32-bit words via the PCI interface and transmitted as two
16-bit words over the link. In addition, each 32-bit word has two additional
out-of-band control bits available which can mark the beginning and end of data
blocks. An overall block diagram of the LTB is shown in Fig.~\ref{fig:ltb}. The
clock frequency for the link transmitter is 64 MHz, double the PCI clock rate.

\subsection{Hotlink Transmitter Card\label{ss-hotlink}} 


Communication between subsystems of the L2 D\O\ trigger must adhere to a common
hardware and software standard.  The protocol uses Cypress Hotlink transmitters
and receivers to provide serial communications operating at 160 Mbs.  A block
of data (typically corresponding to data for one collision) is bounded by
D\O-specific header and trailer words.

Rather than building the Hotlink transmitters as part of the TFC, a separate
transmitter board was built increasing modularity.  The board conforms to the
PC-MIP standard, and one transmitter board is used on each motherboard having a
TFC.  Data are written to the Hotlink transmitter using PCI burst transfers
with the TFC acting as bus master.  The data input to the transmitter must have
a complete D\O\ L2 header and trailer, but the transmitter inserts padding
words needed to meet the D\O\ requirement that the data--block word count be a
multiple of four 32--bit words.  For one TFC/sextant both CTT data and STT fit
data are written.  The CTT data is written as soon as all input data for the
collisions are in the TFC input buffer, and the STT fit data are written as
soon as all fits for a given event are finished.

\begin{figure} 
 \insertfig{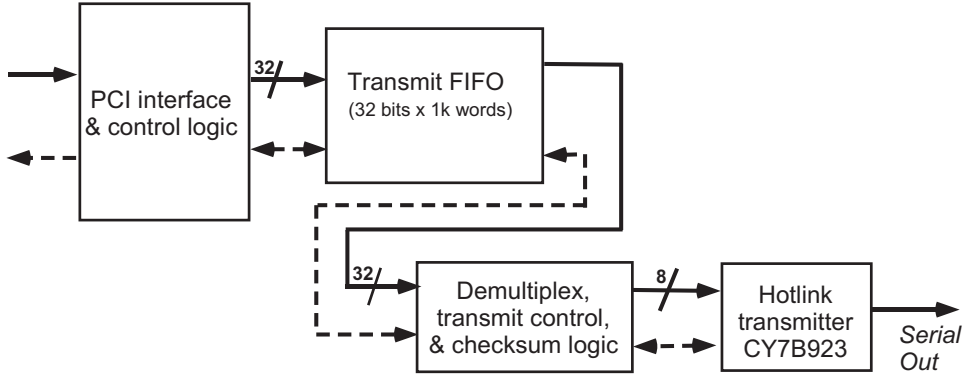} 
 \caption{A block diagram of the Hotlink transmitter and data flow.
   \label{f-hlt}} 
\end{figure}

A block diagram of the Hotlink transmitter is shown in Fig.~\ref{f-hlt}.  Each
transmitter has the oscillator used to drive the Hotlink serial data, a Cypress
Hotlink transmitter~\cite{b-cyhot}, transformers for isolation and an Altera
10k50 FPGA providing the PCI interface, 4 KB of data buffering and Hotlink flow
control.

\subsection{L3 Read Out: Buffer Manager and Buffer Controller\label{ss-l3}}
%

On every L2 accept issued by the trigger system, data corresponding to
the bunch crossing for which the L2 decision was made is read from all
elements of the \Dzero\ detector into the L3 system. The buffering of
this L3 data is differs from sub--systems to sub--system, however, the
read out mechanism is uniform. Single board computers (SBC)
\cite{b-d0b}, housed in each sub-system's VME crates and running the
Linux operating system, are used to gather L3 data from individual
sub-system elements over the VME backplane.
More details on the \Dzero\ L3 trigger system are available in
Ref.~\cite{b-d0b}.

\begin{figure}
  \insertfig{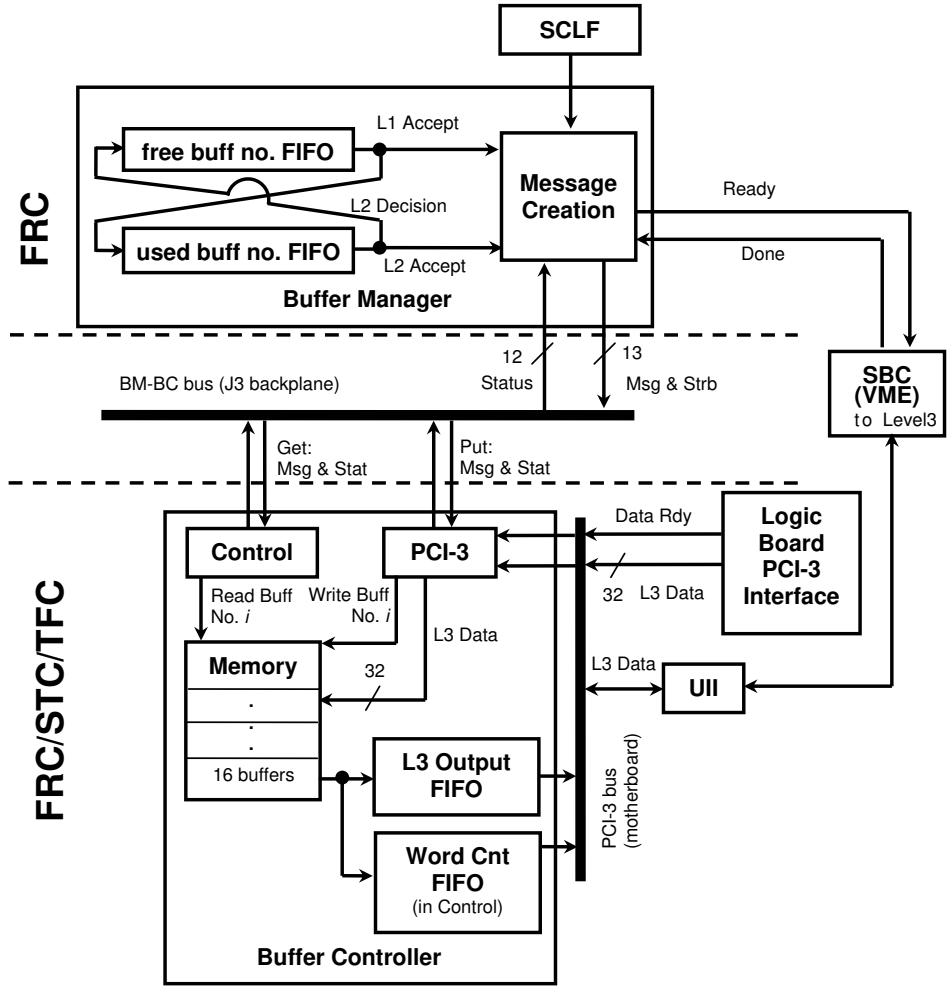}
  \caption{A block diagram of the main functional elements of the
	   STT L3 read out chain. \label{fig:l3blk}}
\end{figure}

Within the STT system, read out to L3, via the SBC, is controlled by
the Buffer Manager (BM) in the FRC. L3 data for each of the individual
boards in the system is stored in custom-built daughter boards called
buffer controllers (BC). The BM broadcasts commands over a group of
dedicated lines on the custom J3 backplane of its crate, telling all
BCs into which buffer they should write data corresponding to an L1
accept or out of which buffer they should read data for transmission
to the SBC. Each BC is responsible for reading data from the logic
board (FRC, STC or TFC) on its motherboard,
storing this data in a buffer at a specified location
and, finally, writing the appropriate data to a VME-accessible FIFO for
read out by the SBC 
when requested to do so by the BM.
The BCs also indicate their progress executing commands to the BM
using status lines on the J3 backplane.
A block diagram of the main components of the STT L3 read out chain is
given in Fig.~\ref{fig:l3blk}.

\subsubsection{Managing the L3 Buffers\label{ss-l3bm}}
Design of the STT L3 read out system is simplified by using identical
memories, divided into 16 fixed-length buffers in all of the BCs. 
In this way, only a single buffer number for
writing or reading needs to be generated in the BM and broadcast to
all BCs.
The BM, then, has three broad tasks to accomplish in managing the L3 buffers
for the STT system:
\begin{enumerate}
  \item Managing lists of used and unused buffer numbers.
  \item Broadcasting the appropriate buffer number to  the BCs for
	writing/reading on L1/L2 accepts
	(see section~\ref{ss-l3comm}).
  \item Notifying the SBC that all L3 data in this STT crate is ready
	to be read out
	(see section~\ref{ss-l3sbc}).
\end{enumerate}

The scheme for allocating and deallocating buffer numbers uses two
FIFOs: a ``free'' FIFO containing a list of buffer numbers that are
available for writing and a ``used'' FIFO with a list of buffer
numbers currently containing data.
On an L1 accept, the next buffer number (if available) is moved from
the free to the used list and is sent to the BCs. L2 decisions are
sequential at \Dzero , so on each L2 period, the next buffer number is
taken from the used to the free list. If the L2 decision is accept,
this buffer number is also sent to the BCs, otherwise no further
action is taken.

\subsubsection{Buffer Controllers\label{ss-l3bc}}
Storage of L3 data, for each L1 accept, pending read out to L3 if an L2
accept is issued, is accomplished in the STT system using a common set
of daughter boards -- the buffer controllers (BC). A BC is plugged into
each motherboard and communicates with the logic daughter board on that
motherboard by the PCI-3 bus. The BCs have four main hardware
elements (see Fig.~\ref{fig:l3blk}):
\begin{enumerate}
  \item L3 data memory divided into sixteen 8K $\times$ 32-bit buffers and
	implemented using four IDT70V9099 128K $\times$ 8-bit synchronous
	dual-port SRAMs.
  \item An L3 output data FIFO used to buffer data for transmission to
	the SBC and implemented using an IDT72V36100 64K $\times$ 36-bit FIFO
  \item An interface to the PCI-3 bus containing both PCI master and
	target functionality used for reading L3 data from the logic
	daughter boards and for the transfer of data to the SBC via a
	PCI to VME interface, implemented in an Altera Flex 10K50
	FPGA.
  \item Control logic for writing and reading data to the memory and
	FIFO, implemented in an Altera Flex 10K30 FPGA.
\end{enumerate}

Upon receipt of a {\it Put} message from the BM on an L1 accept 
(see section
\ref{ss-l3comm}) the BC begins the process of transferring data from
its associated logic daughter board to a buffer specified by the
BM. The logic board to BC transfer starts when the logic board
sets one of the user-defined lines on the PCI bus. To minimize
transfer time, this transaction is done as a PCI burst. However, since
the BC does not know, {a priori}, the number of data words to
transfer, the transaction is terminated by the logic board using the
PCI target disconnect function. As data words arrive from the PCI bus,
they are stored sequentially into the memory block associated with the
chosen buffer number.

If the L2 decision associated with the L1 accept whose data is stored
in buffer {\it i} is a reject, no action is taken at the BC. Buffer
{\it i} is simply returned to the list of free buffers in the BM and
any data in it will be overwritten the next time that buffer is
used. If the L2 decision is accept, however, the BM sends
a {\it Get} message to the BC 
(see section~\ref{ss-l3comm}).
The BC control logic then transfers data from the buffer number
specified in the BM message to the L3 output FIFO. It also stores the
number of words in the data block in another FIFO, {\it word count},
implemented in the PCI-3 interface FPGA. Data wait in 
these FIFOs until they are read by the SBC through the PCI-to-VME bridge on
the motherboard. The SBC first reads the word count FIFO, as well as
crate identifier information, and then performs a block transfer from
the L3 output FIFO of the number of words specified by the word count.

\subsubsection{BM-BC Communication Protocol\label{ss-l3comm}}
Messages are passed between the BM and the BCs using a 
25-line bus on the
custom J3 backplane. The protocol is very similar to that used for
messaging between Fermilab VME Read Out Buffers (VRB) and the VRB
Controller (VRBC) \cite{VRB}.
Messages from the BM to the BCs occupy 12 lines, 
four of these are used to identify the message type
while the remaining eight contain data associated with the
message. 
One line is used as a message strobe.
Six status signal are passed as levels from the BCs to the BM.
Each status signal is sent as a differential pair on two lines,
which are ORed or ANDed on the backplane. 
The BM therefore receives the combination of the status from all BCs.

Messages sent from the BM to the BCs come in two classes.
\begin{enumerate}
  \item {\it Put} messages, indicating that the BCs should read L3 data
	from their associated logic boards and put that data into the
	buffer. 
  \item {\it Get} messages, indicating that the BCs should read data
	from their buffer and write it into the L3 output FIFO for
	read out (described below) by the SBC.
\end{enumerate}
The protocol for {\it Put} and {\it Get} classes is identical. Each
class requires three separate messages and one status signal from the
BCs. To begin the {\it Put} or {\it Get} process, the BM sends two
messages.
The first contains the buffer number to be used for the write or read
process, while a bunch crossing identifier is sent with the second and
is used in the BCs to check for event misalignment.
When a BC has finished its {\it Put} or {\it Get} operation, it sets
its input to the {\it Done} status line. 
The {\it Done} and {\it Done*} signals from all BCs are ANDed and
ORed on the backplane and indicate to the BM when the first BC, and
when all BCs, have finished their operations.
When all BCs are done, the BM sends out its final message of the
chain to indicate that the entire {\it Put} or {\it Get} process is
finished. 

Besides the {\it Done} status lines, the BCs also transmit error and
busy information from their associated logic boards to the BM. 
Based on these status signals the BM can request that the trigger
framework issue an initialization command in the case of a serious
error, or that it block L1 accepts when buffers in the STT system
become full due to processing delays.

\subsubsection{Communication with the SBC\label{ss-l3sbc}}
The BM notifies the SBC that all L3 data are ready for read out after an
L2 accept using a user-defined line on the VME J2 backplane. The SBC
then takes over the process of reading all of the L3 data from its STT
crate as described in Ref.~\cite{b-d0b}. When read out is finished, the SBC
notifies the BM using another user-defined line on the VME J2 backplane.

\section{Silicon Track Trigger: Simulation\label{s-trigsim}}
As with other D\O\ trigger elements, a full simulation of the STT was
developed.  The simulation was used in the design phase of the STT for
algorithm testing, is used with simulated signal and background samples to test
the selection efficiency for different trigger criteria, is used for
verification of results determined using the actual hardware, and it contains
the code to generate most of the hardware look up tables.

The full D\O\ trigger simulation (trigsim) is a single program which provides a
standard framework for including code for individual trigger elements.  The
core pieces of the framework give a means to specify the data format
transferred between trigger elements, a means to simulate the time ordering of
the trigger levels and a simulation of the data transfers.  Algorithms specific
to a given (hardware) trigger element are required to receive and send data in
the actual format used online using an interface provided by the trigsim
framework.  The trigger simulator code is written in C++, although for part of
the STT simulation, the C++ is used to redirect calls to standard C code.

The STT simulation has three major elements motivated by the hardware design.
These are: (1) an emulation of the FRC functionality, (2) an emulation of the
STC functionality and (3) emulation of the TFC which can use the actual C code
run in the DSPs.  In addition, C++ classes were developed to represent the
input and output data, including expanded formats used online for debugging
purposes.  The internal structure of the simulator mimics the sextant and
sector division of the actual STT. As for all trigger elements in the
simulation, the STT code also creates output data formatted as sent to L2 and
L3 online.  The STT simulation can be run either as part of the full D\O\
trigger simulation or in a standalone mode.  The standalone mode is used
primarily for testing and debugging.

In the FRC hardware, algorithms are implemented using programmable logic chips.
Because of this, the trigger simulator provides an emulation of the FRC
functionality written in C++.  All major features of the FRC are provided,
including subdivision into six sextants and redistribution of the (simulated)
inputs from the L1CTT.

Like the FRC, the STC hardware uses programmable logic arrays and the simulator
also implements the algorithms via emulation using C++ code.  Historically, the
cluster-finding algorithm was developed in C++ and then translated into the
firmware running on the STC.  Many of the look up tables used in the STC
hardware, for example the tables which determine the boundaries of the
cluster-matching region for each L1CTT track, can be created within the
simulator.  The look up tables generated within trigsim can be written out in
the format used online, and the actual online look up tables can also be read
back into the simulator.  This provides an efficient means for generating and
studying the online performance.

The TFC hardware is built using a combination of programmable logic and
standard DSPs programmed via C code.  The simulator therefore uses a mixture of
emulation (for the event building done in programmable logic) and the actual
fitting code.  As with the cluster finding for the STC, the track-fitting
algorithms used online in the DSPs were first developed in C++.  The first
algorithms were developed with floating-point precision within the framework of
the STT simulator.  Several different algorithms were studied using Monte Carlo
simulated D\O\ data to determine their efficiency for selecting long-lived
tracks and their rejection of prompt events.  The floating-point algorithms
were then translated into algorithms using only integers.  These integer
algorithms were first implemented in the simulator, and then, in the final step,
the actual C code for the DSPs was written.  The simulator includes not only
the capability of using the prototype floating point and integer routines, but
also can have the C code for the DSPs compiled in.  The precomputed lookup
tables used in the TFC hardware for the track fitting, described in
Section~4.4.2, are generated within the STT simulation, and the simulation can
read these back in for use with the DSP code in the simulator.

The STT simulation also provides a mechanism to produce test vectors, which can
then be input into the hardware for local and global STT testing.  The output
test vectors can be compared directly to the resulting hardware output
bit--by--bit.  These test vectors are used to verify algorithms, hardware
configuration and hardware operation {in situ}.

The STT simulation also runs on real D\O\ data.  Once the STT hardware data was
written into the D\O\ data stream, it became possible to run the STT simulation
for either the actual simulation or in {\em pass-through} mode, whereby the STT
data sent to L3 is analyzed directly.  Both modes of operation were used for
verifying the STT output data from the hardware.  In addition, the simulation
can operate in mixed modes in which some of the data are passed through as
computed online and some are regenerated in the simulation.  For example, the
L1CTT tracks found online can be used as input to either the STC simulation, or
the TFC simulation, or both. 

\section{System Performance\label{s-performance}} 


The STT performance has two aspects.  One is the overall processing time in
comparison with the allowed time budget.  The other is how well the physics
goals are met.  The second of these depends on a number of variables including
CTT road multiplicity, SMT cluster reconstruction, pattern recognition
performance and track fitting quality. This section describes results for both
aspects of the performance.

\subsection{STT Processing Time\label{ss-proctime}}
The general D\O\ L2 processing budget determined from queuing simulations is
100~$\mu$s total for processing by the preprocessors and L2 global, with
another 100~$\mu$s allowed for latency.  These simulations, which assume
buffers for 16 events at all time critical points in the processing, were used
as guidelines for the STT design.  The actual STT processing time for each
interaction depends on the number of CTT tracks, the occupancy of the SMT
detector, and the number of clusters that get assigned to CTT roads.  There is
also a dependence on initial STT track--fit quality.  All of these quantities
are affected by the instantaneous luminosity, and these determine the
variation in actual processing times, the data transfer times and the latency.

The STT processing occurs with a high degree of parallel computation.  For
example, the SMT data are converted to clusters on-the-fly, and the time for
the last cluster to be found is essentially the same as the read out time for
the last of the data from the SMT.  Similarly, the CTT data input to the FRC is
rebroadcast to the STCs and TFCs within a few clock cycles.  Thus, most of
the processing by STT appears as latency, and is not counted against the
processing time budget.

\begin{figure}
\insertyfig{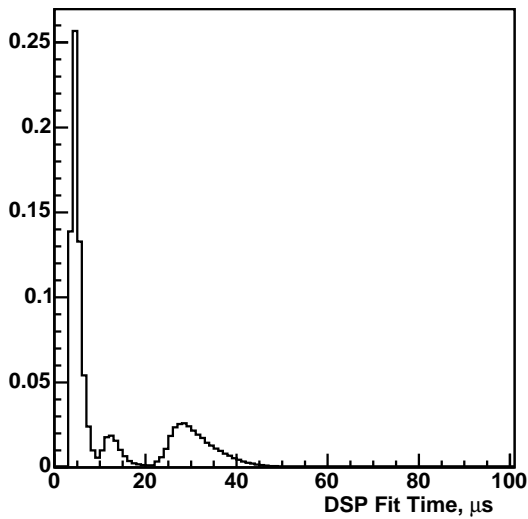}{3truein}
\caption{The DSP processing time per fit distribution. This is the
  time from when the fit data are loaded into the DSP until the DSP
  has written results to the output buffer memory.  The first spike
  occurs for fits in which there are too few hits for a fit to be
  performed.  The small bump is for single pass fits, and the second
  larger bump is for two pass fits.  The fraction of fits in each
  feature is accelerator condition, detector condition and luminosity 
  dependent.  The plots shown here were taken at an instantaneous luminosity 
  of $30\times 10^{30}$/cm$^2$/s, and the mean fitting time is $14\ \mu$s/road.
  \label{f-tfctime}}
\end{figure}

The main exception to this is the track fitting performed in the TFC.  Although
there is significant parallel capacity, an individual fit can take up to
50~$\mu$s (although most occur much faster).  During this time, if other DSPs
in the same half-TFC are free, additional roads can be fit.  However, if all
DSPs are used, then fit input data waits in buffers until a DSP is available. 
Fig.~\ref{f-tfctime} shows the time taken to load data into a TFC from the
internal TFC buffer (IDPM) until the fit output data are written to the output
buffer. 

The overall STT processing time and latency do fit well within the allowed
budget.  During D\O\ data taking, at even the highest instantaneous
luminosities seen thus far, the STT contributes a negligible amount
to overall experimental dead time.

\subsection{Reconstruction and Physics Performance\label{ss-quality}}
The overall effectiveness of the STT will be determined by the signal
efficiency and background rejection that it delivers.  This is largely
determined by the fidelity of the reconstructed tracks which, in turn, depends
on the quality of the SMT cluster reconstruction and the pattern recognition
and fitting algorithms.

\begin{figure}
\insertfig{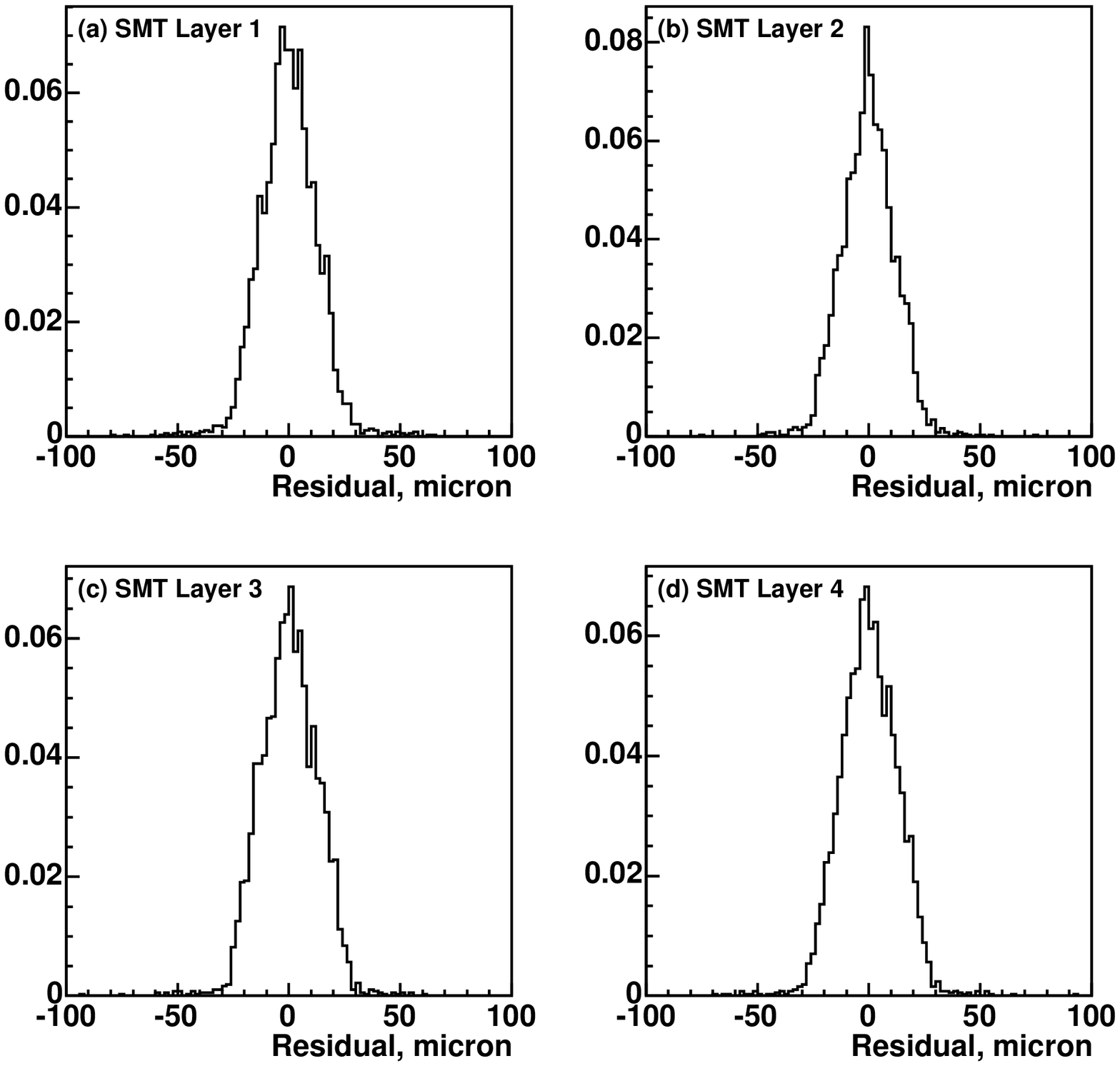}
\caption{The distance between the reconstructed STT cluster position and
 the trajectory of the matched particle in each of the four SMT layers
 for a simulated sample of $p_T=50$~GeV/$c$ single muons.
 As with the STT fitting algorithm, the trajectory is assumed to be circular; 
 no multiple scattering effects are considered.  \label{f-cluspos-mu50}}
\end{figure}

\begin{figure}
\insertfig{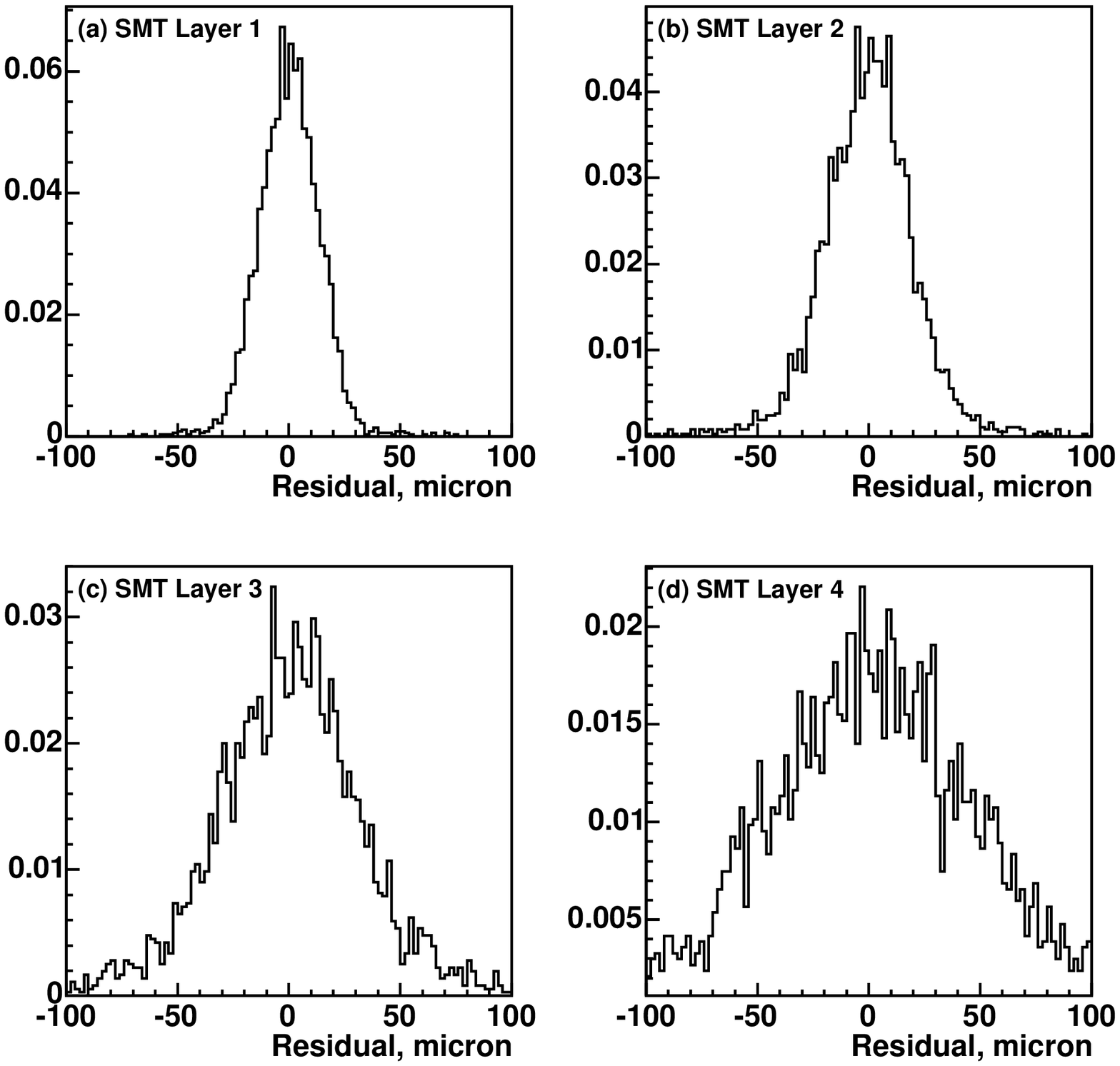}
\caption{The distance between the reconstructed STT cluster position and
 the trajectory of the matched particle in each of the four SMT layers
 for a simulated sample of $p_T=2$~GeV/$c$ single muons.
 As with the STT fitting algorithm, the trajectory is assumed to be circular; 
 no multiple scattering effects are considered.  \label{f-cluspos-mu02}}
\end{figure}

\begin{figure}
\insertfig{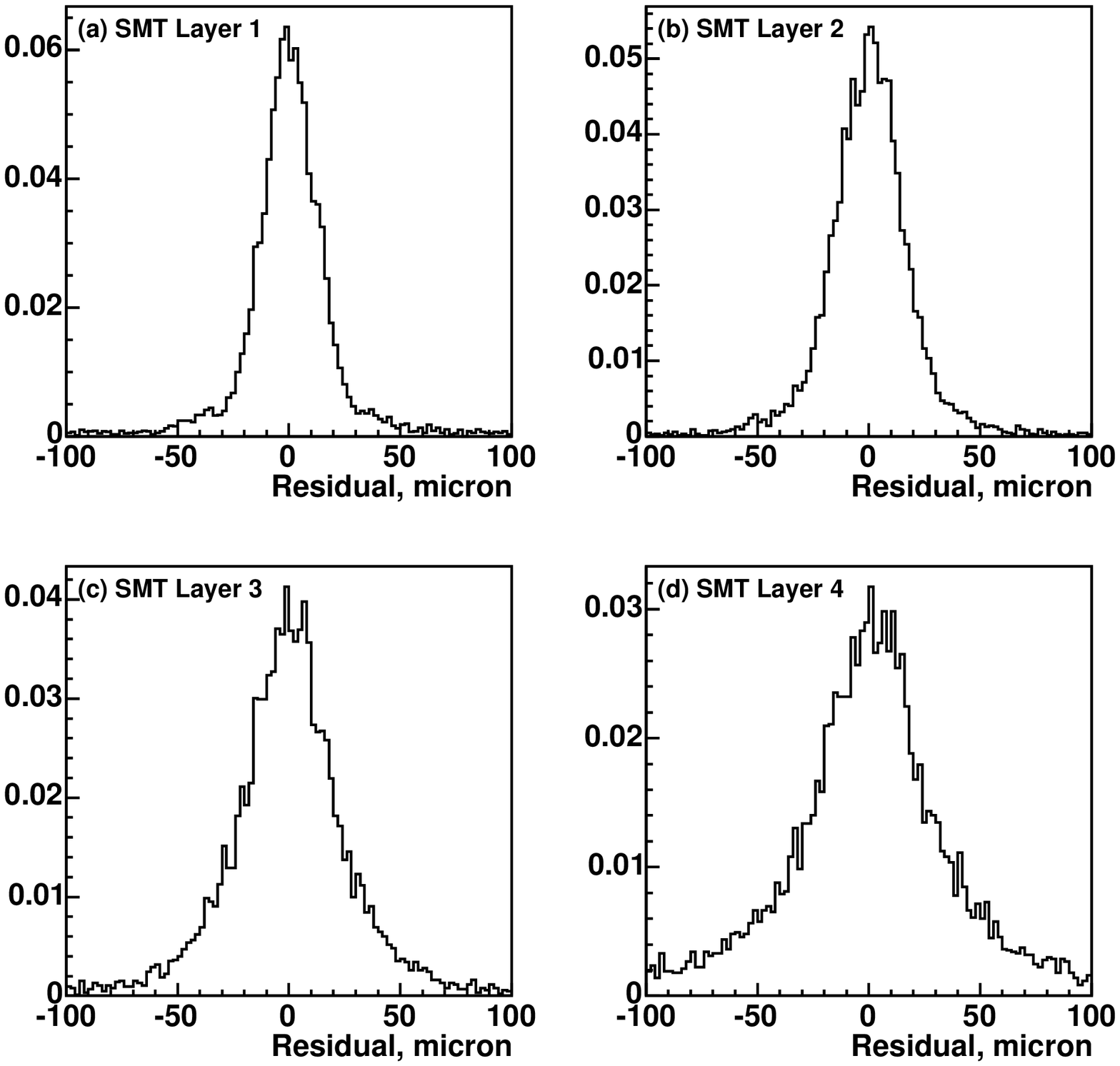}
\caption{The distance between the reconstructed STT cluster position and
 the trajectory of the matched particle in each of the four SMT layers
 for a simulated sample of $ZH\rightarrow\nnbar\bbbar$ events.
 As with the STT fitting algorithm, the trajectory is assumed to be circular; 
 no multiple scattering effects are considered.  \label{f-cluspos-zh}}
\end{figure}

  Fig.~\ref{f-cluspos-mu50} and~\ref{f-cluspos-mu02} show the $r\phi$
position difference between true particle trajectories at each SMT layer and
the associated SMT clusters at the same layer.  The data are simulated single
muons of $p_T=50$~GeV/$c$ and $p_T=2$~GeV/$c$ respectively.  One sees clearly
the effect of multiple scattering in the lower $p_T$ sample.  The quality of
the association is dependent on the physics as well.  Fig.~\ref{f-cluspos-zh}
shows the same distributions for tracks from simulated
$ZH\rightarrow\nnbar\bbbar$ reactions.  These distributions are created from
all good quality reconstructed tracks.  A good quality track is defined as one
which satisfies\footnote{This was defined using simulated single
muon events.  The value used for collider data is somewhat looser; the
coefficient on the square root is typically 10.} $\chi^2 \le 4\sqrt{1 + (4\ 
\mathrm{(GeV/c)}/p_T)^2}$.  The $p_T$ dependence reflects
the effect from multiple scattering.

\begin{figure}
\insertfig{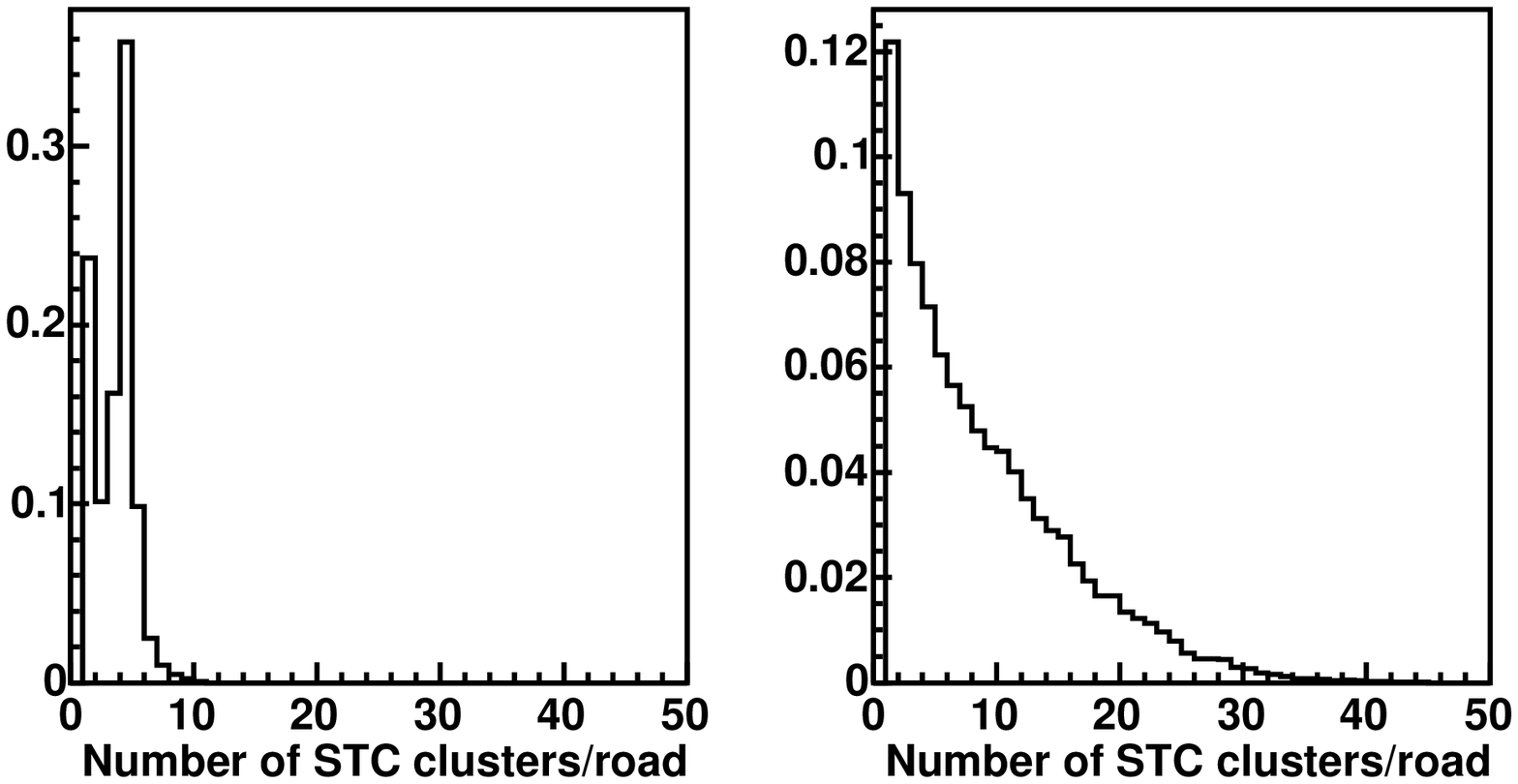}
\caption{The number of STT clusters/CTT track in (a) simulated single muon 
events and (b) simulated $ZH$ events.\label{f-nclus}}
\end{figure}

\begin{figure}
\insertfig{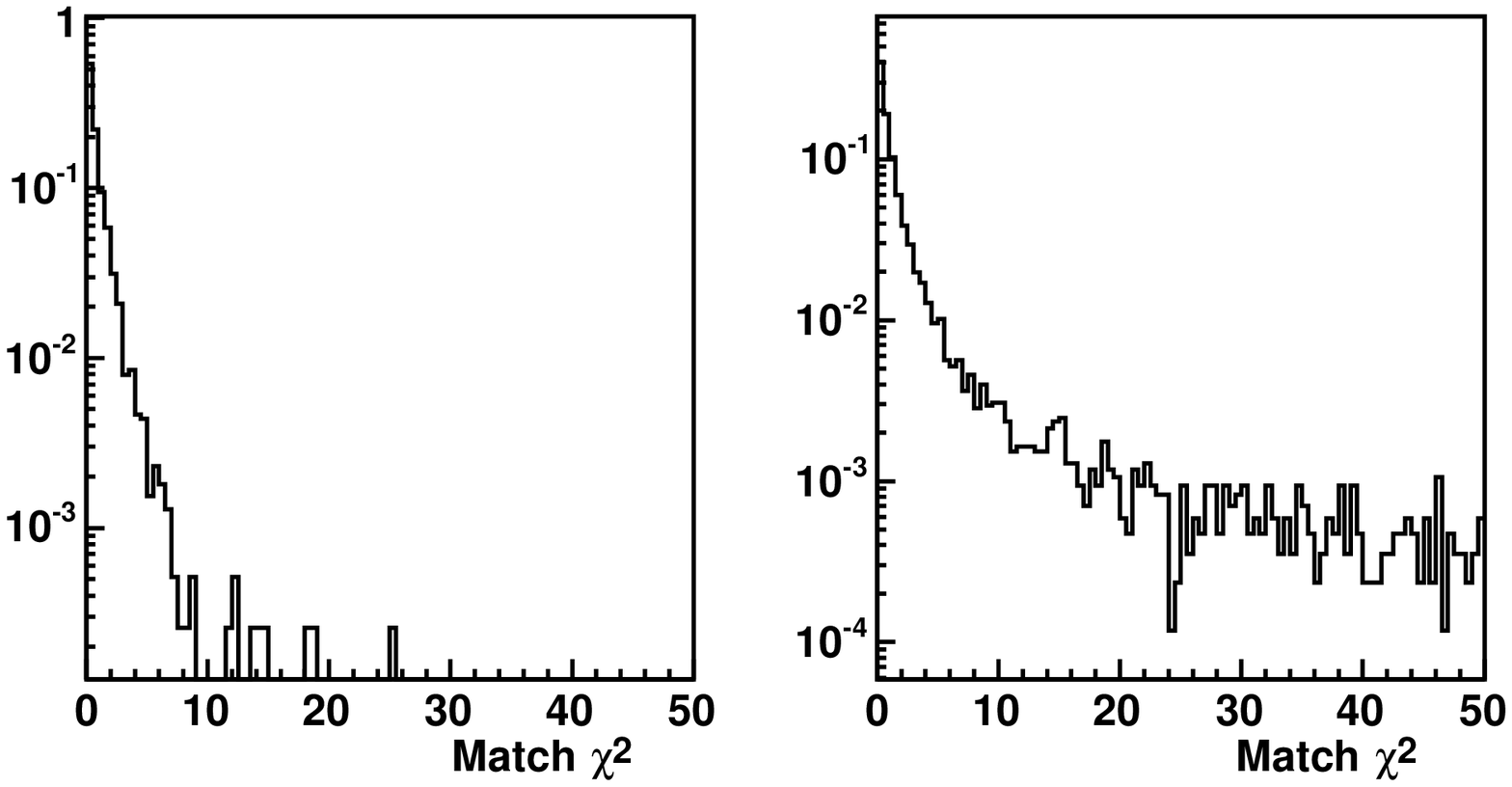}
\caption{The match $\chi^2$ obtained from the difference in track parameters
$\delta b,\ \delta\phi$ and $\delta\kappa$ between a reconstructed STT track  
and the particle trajectory that gives the best $\chi^2$.  The left--hand
panel is for simulated single muons, and the right--hand panel is for
simulated $ZH$ events. \label{f-fitmatch}}
\end{figure}

Fig.~\ref{f-nclus} shows the cluster multiplicity per road after the initial
filtering for simulated single muon and $ZH$ interactions.  One
sees the relative cleanliness of the simple single muon sample as compared with
the $ZH$ physics sample, and the reason for the final filtering pass.
Fig.~\ref{f-fitmatch} shows the distribution of the $\chi^2$ calculated from
the difference between fit parameters for a fitted track and its matched true
Monte Carlo particle.  The fit error matrix is used in the calculation.  The
matched particle is determined by trying all possible matches between a given
fit track and all true particles with $p_T>1.5$~GeV/$c$.  The particle which
gives the lowest $\chi^2$ is defined as the matching particle.  There are 3
degrees of freedom in this $\chi^2$, and one sees reasonable match $\chi^2$
values.

\begin{figure}
\insertfig{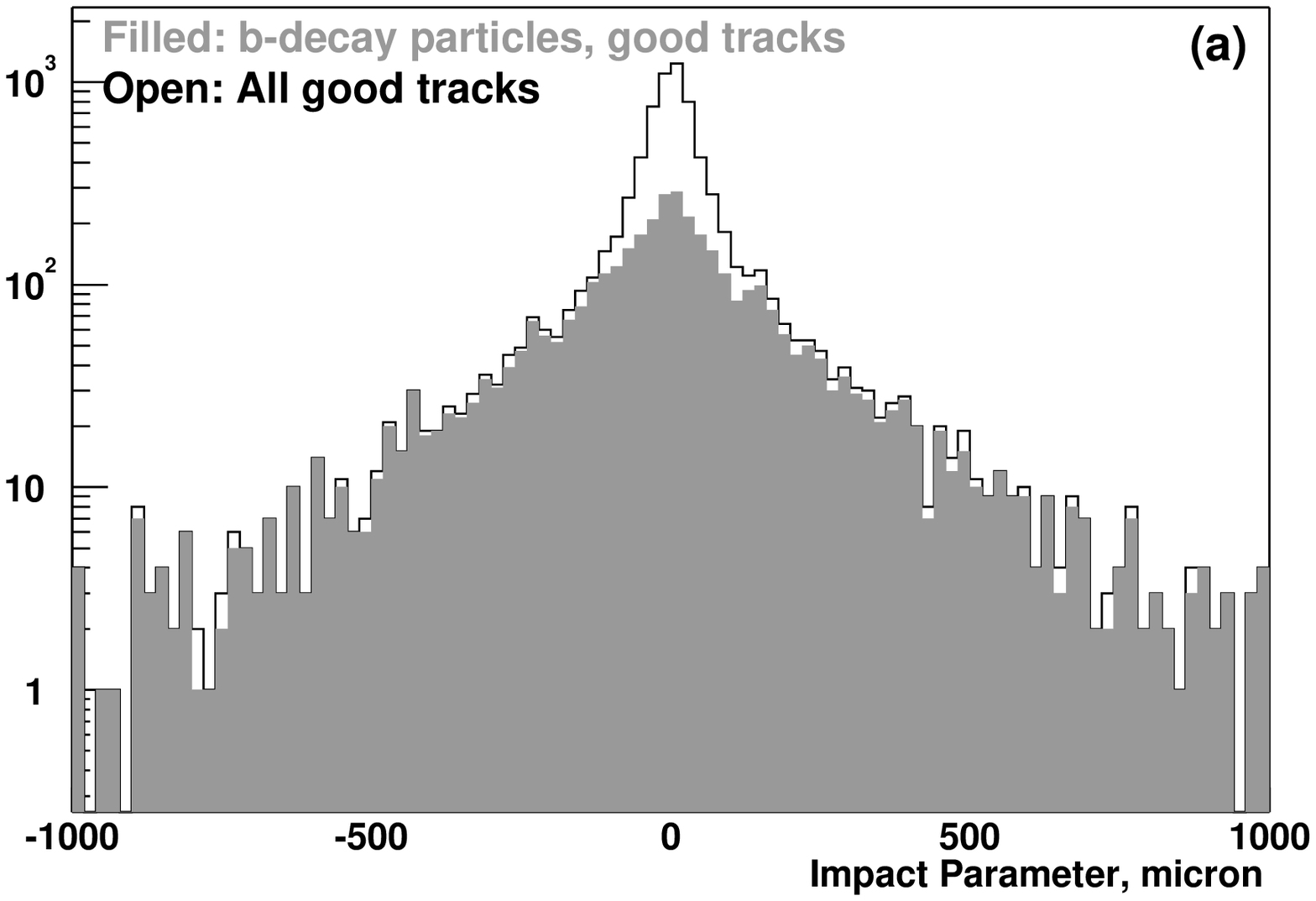}\insertfig{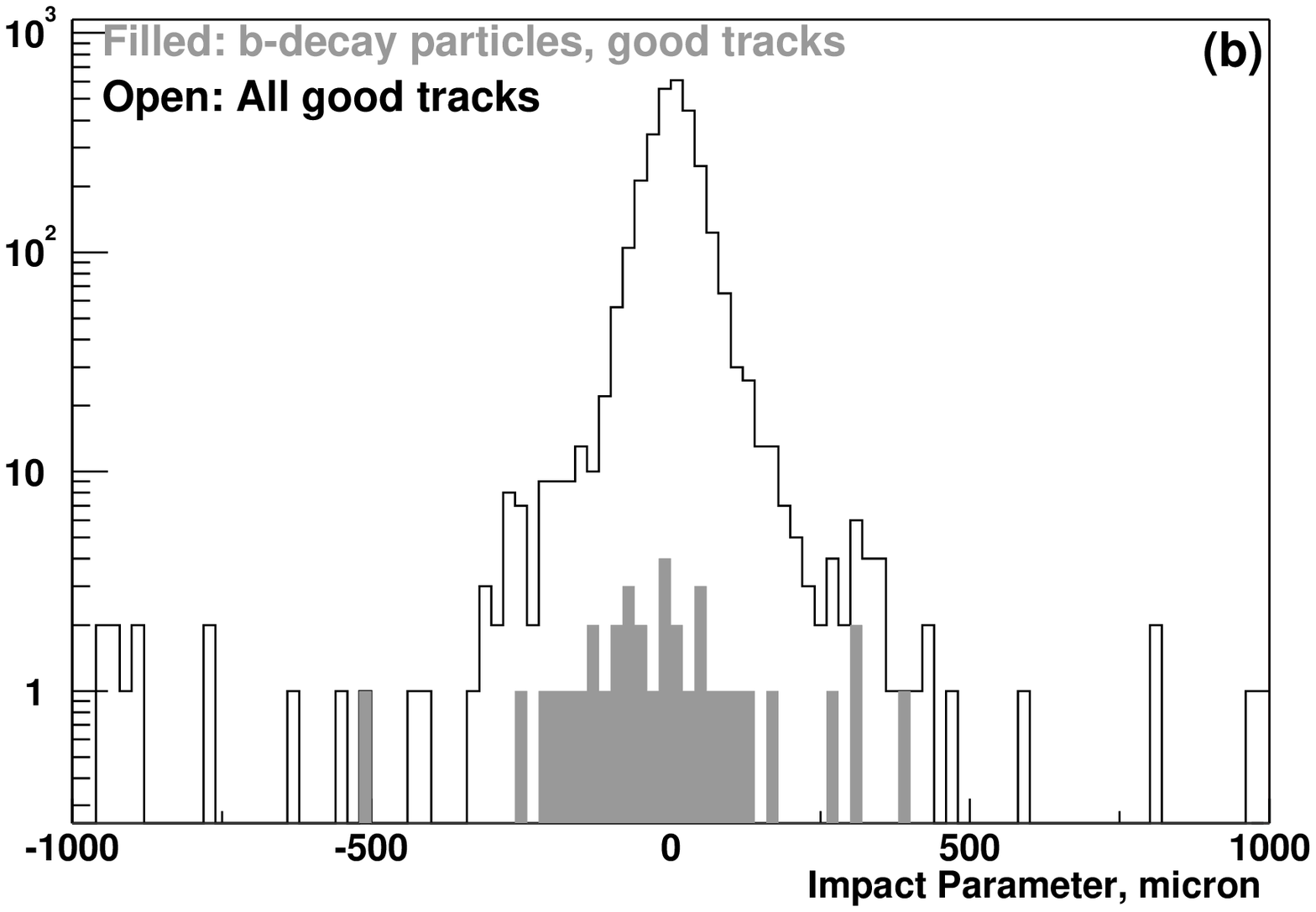}
\caption{The reconstructed impact parameter distributions for (a) 
 simulated $ZH$ events and (b) simulated QCD events. \label{f-bimp}}
\end{figure}

\begin{figure}
\insertfig{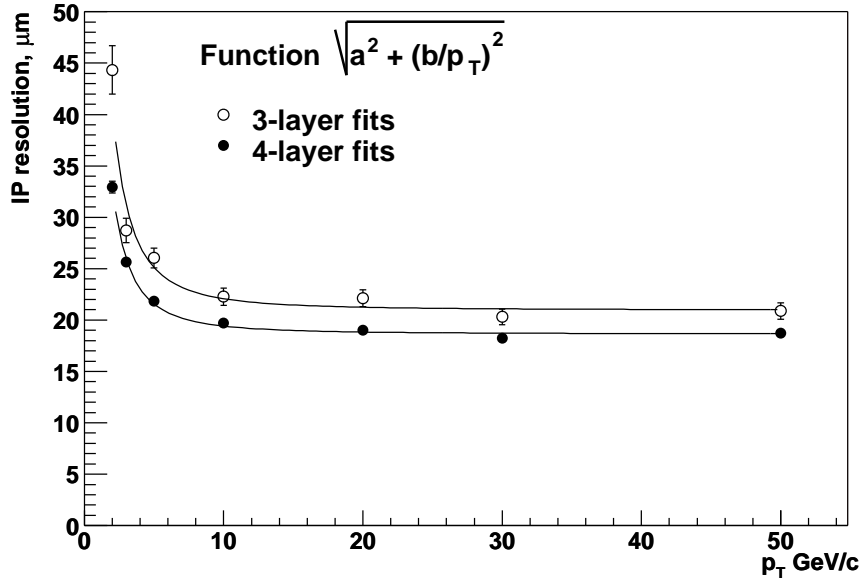}
\caption{Impact parameter resolution as a function of $p_T$ as determined
  from simulated single muon events.  The upper curve is for fits which
  have hits from three SMT layers; the lower, for fits with four. The
  parameter values for the upper curve are $a = 21\pm0.4\ \mu\mathrm{m},
  \ b=69\pm3\  \mu\mathrm{m/GeV/c}$. For the lower curve, the values are 
  $a = 18.6\pm0.1,\ b = 54\pm1$ with the same units.
  \label{f-mures}}
\end{figure}

\begin{figure}
\insertfig{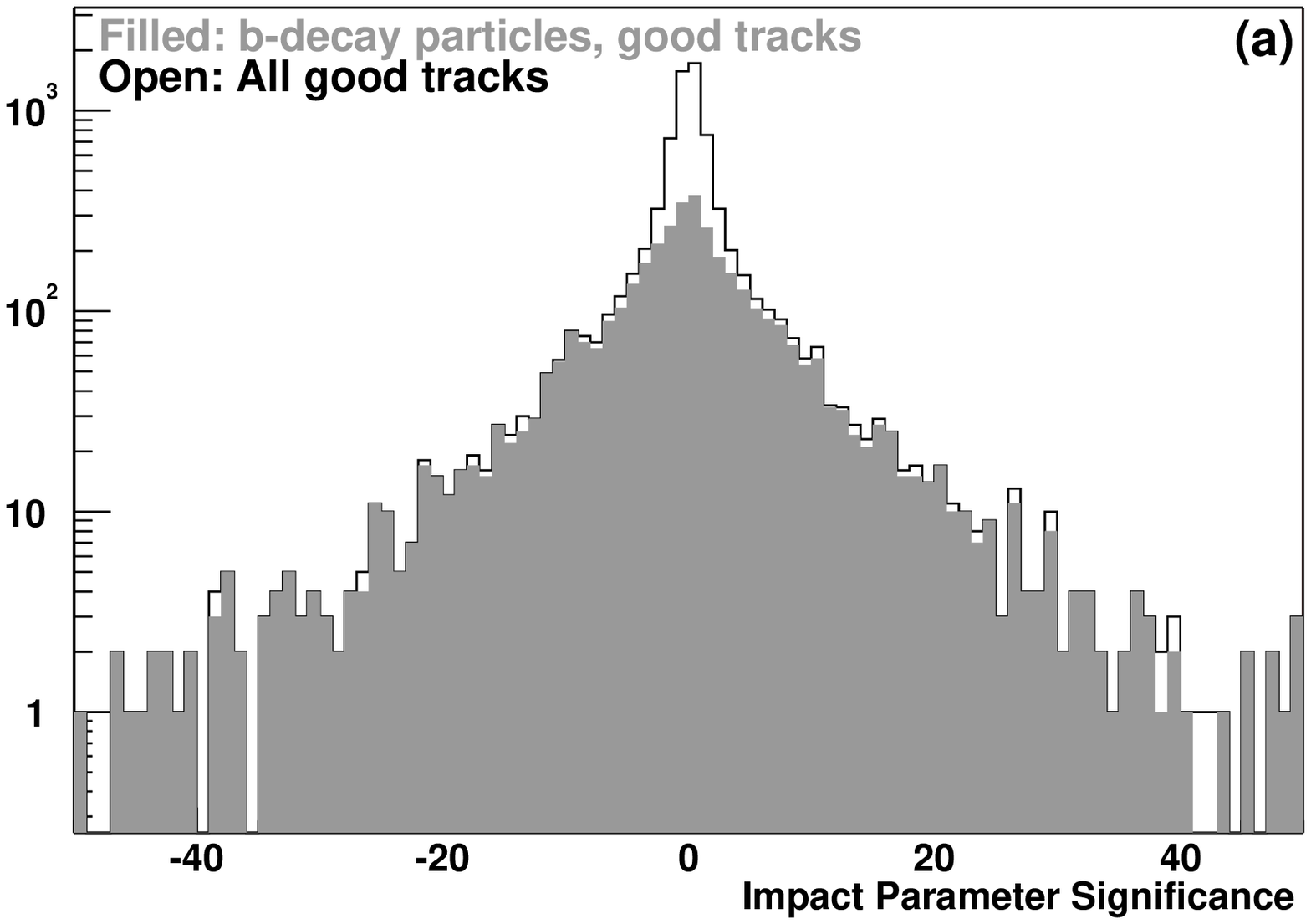}\insertfig{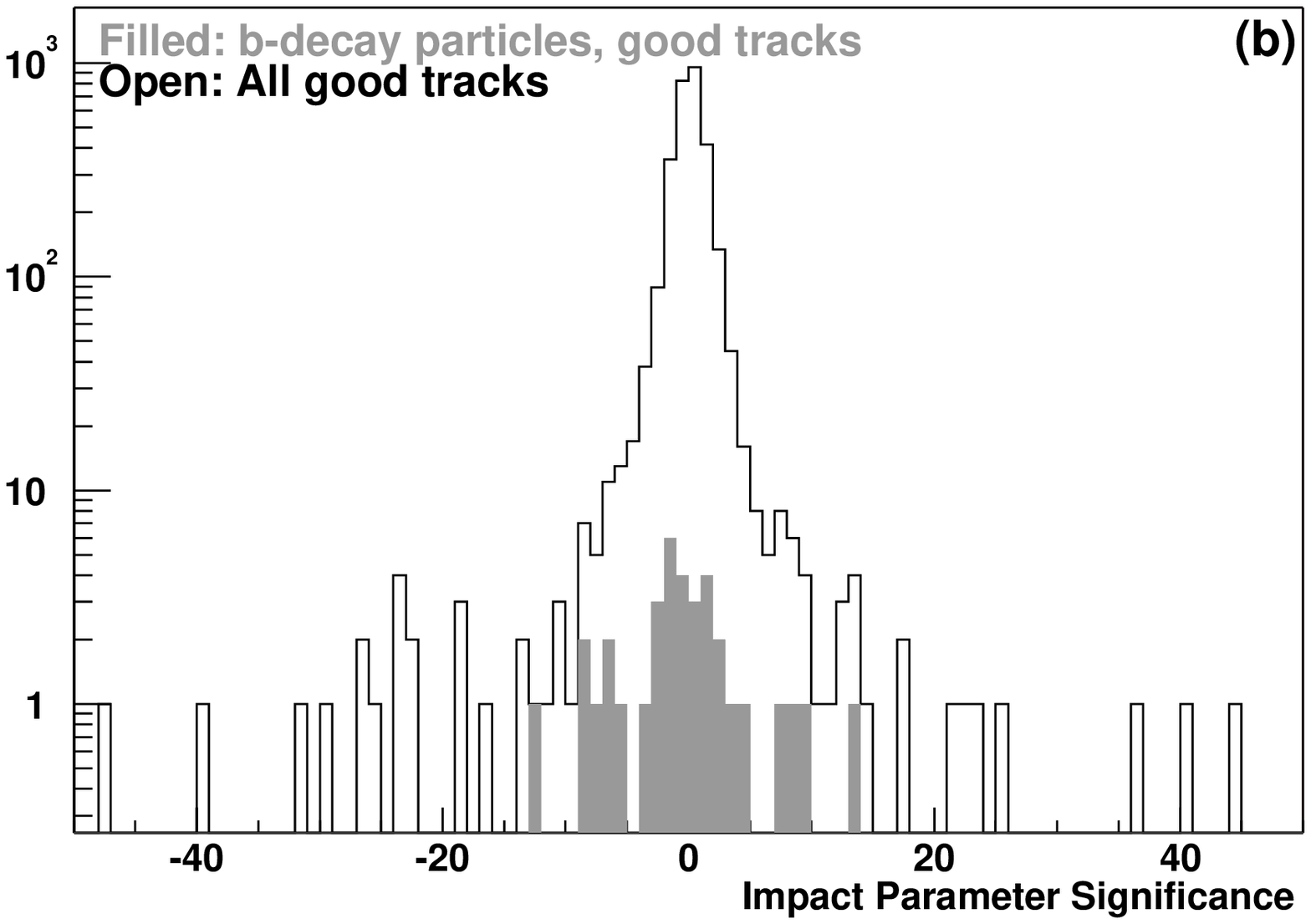}
\caption{The reconstructed impact parameter significance distributions for (a) 
 simulated $ZH$ events and (b) simulated QCD events. \label{f-bsig}}
\end{figure}

Fig.~\ref{f-bimp} shows distributions of the reconstructed impact parameter
for all good STT tracks in two samples: (a) simulated $ZH$ events and (b)
simulated dijet events.  The shaded region corresponds to good tracks for which
the matched true particle comes from a $b$--flavored decay.  The open histogram
is for all good tracks.  One sees clearly the predominance of $b$--flavor in
the large impact parameter regions.  The impact parameter distribution width
has a $p_T$ dependence introduced by multiple scattering.  The effect of the
$p_T$ dependence can be reduced by using the impact parameter significance $S_b
\equiv b/\sigma_b$ instead of the impact parameter.  The
uncertainty\footnote{This value was determined by fitting the width seen for
single muons as a function of $p_T$ using tracks with four SMT clusters.}
$\sigma_b = \sqrt{a^2 + (b/p_T)^2}$ takes into account the effect of multiple
scattering.  Fig.~\ref{f-mures} shows the impact parameter resolution
obtained from simulated single muons.  The lower curve is for the case when a
fit used four SMT clusters, and the upper curve is for the case when a fit used
only three SMT clusters.  Fits to these curves give $a = 18.6\pm0.1\ \mu$m and
$b=54\pm1\ \mu$m/(GeV/$c$) for fits with four SMT clusters and $a = 21.0\pm0.4\ 
\mu$m and $b=69\pm3\ \mu$m/(GeV/$c$) for fits with three SMT clusters.
Fig.~\ref{f-bsig} shows the impact parameter significance distributions for
the same two samples in Fig.~\ref{f-bimp}.  Here, the separation between
signal and background is even clearer than for the impact parameter alone.

The STT tracks can be used as input to a fast multivariate b-identification
algorithm which is run in the L2 global processor.  The algorithm combines
various track quantities in a way which enhances the online b-identification
efficiency compared to simpler IP cut methods.



\begin{figure}
\insertfig{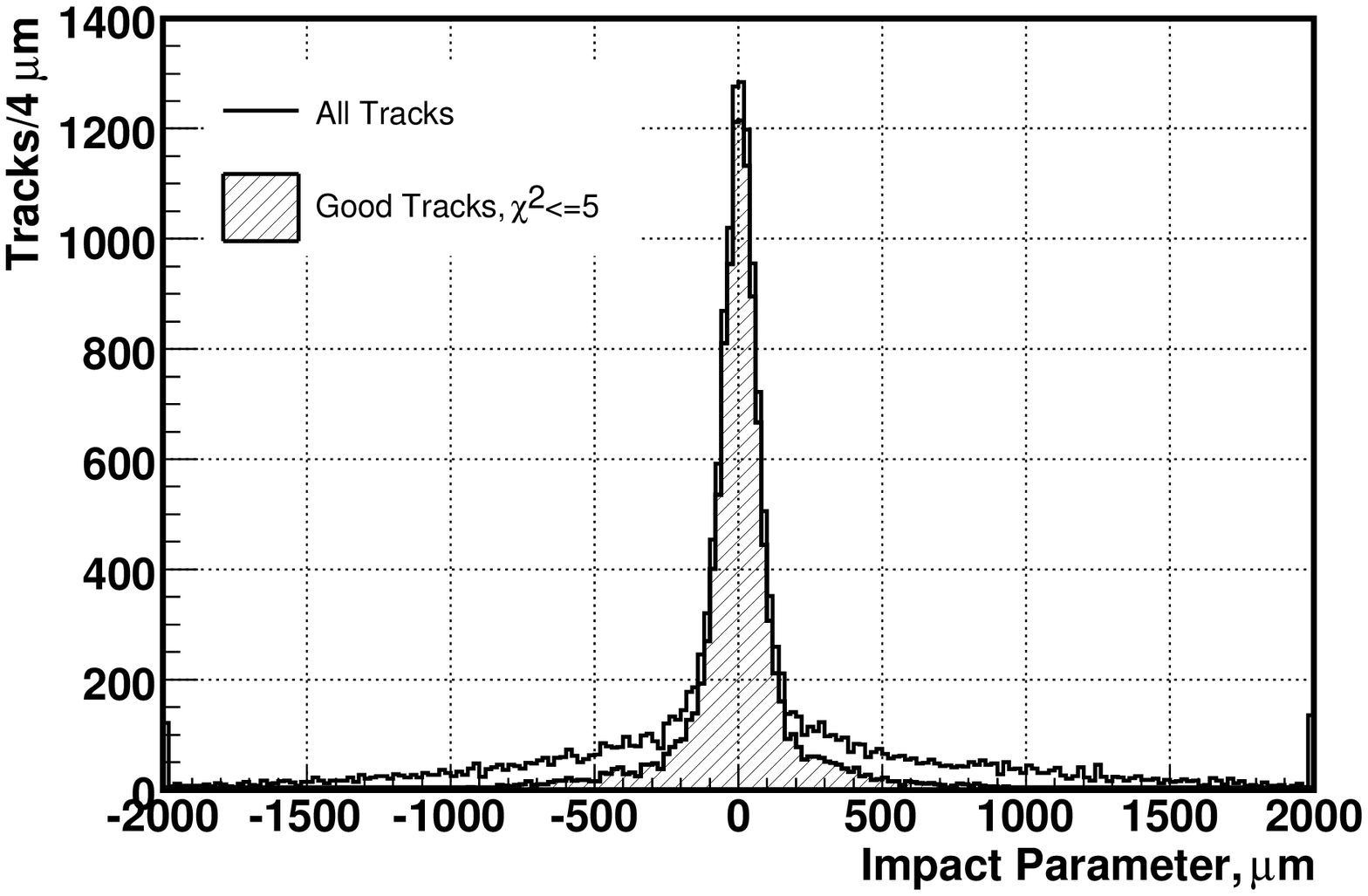}
\caption{The reconstructed impact parameter distributions for tracks in 
 collider data.  No trigger selection is applied, so most events are multijet
 events with relatively low momentum tracks.\label{f-ip}}
\end{figure}

Fig.~\ref{f-ip} shows the impact parameter distribution reconstructed by the
STT in collider data taken at an instantaneous luminosity of roughly $15\times
10^{30}$/cm/s.  The open histogram shows the result for all tracks
reconstructed by the STT, and the hatched histogram shows the result for good
tracks, defined by requiring the fit $\chi^2$ to satisfy 
$\chi^2/\sqrt{ 4 + (8/p_T)^2} \le 5.0$.  One
sees the good track requirement has very little impact on the efficiency but
removes tails from misreconstructed tracks.

\begin{figure}
\insertfig{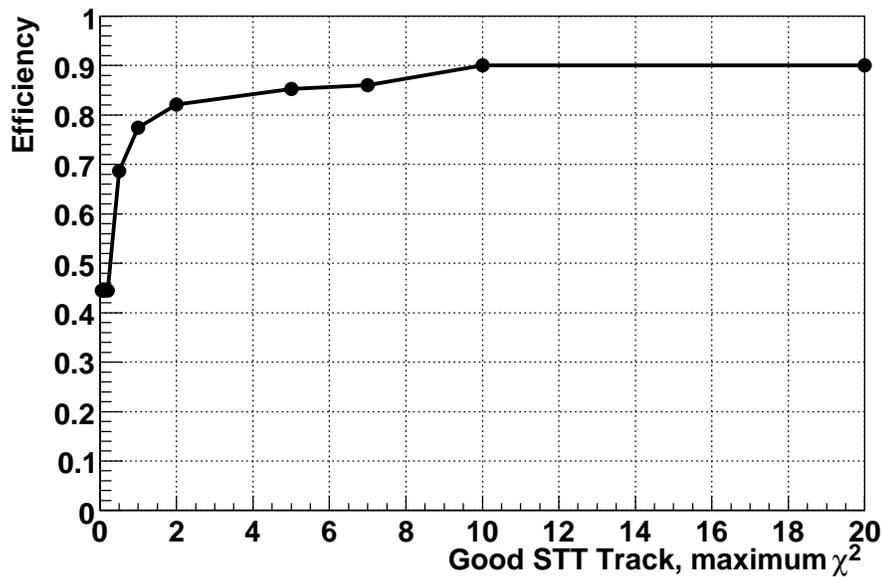}
\caption{The STT track reconstruction efficiency relative 
 to the offline reconstruction as measured using muons in $Z\rightarrow\mu\mu$ 
 decay. The horizontal axis is the maximum fit $\chi^2$ allowed for a ``good'' 
 STT track.  The statistical uncertainties are too small to be seen on this
 scale.
 \label{f-zeffi}}
\end{figure}

\begin{figure}
\insertfig{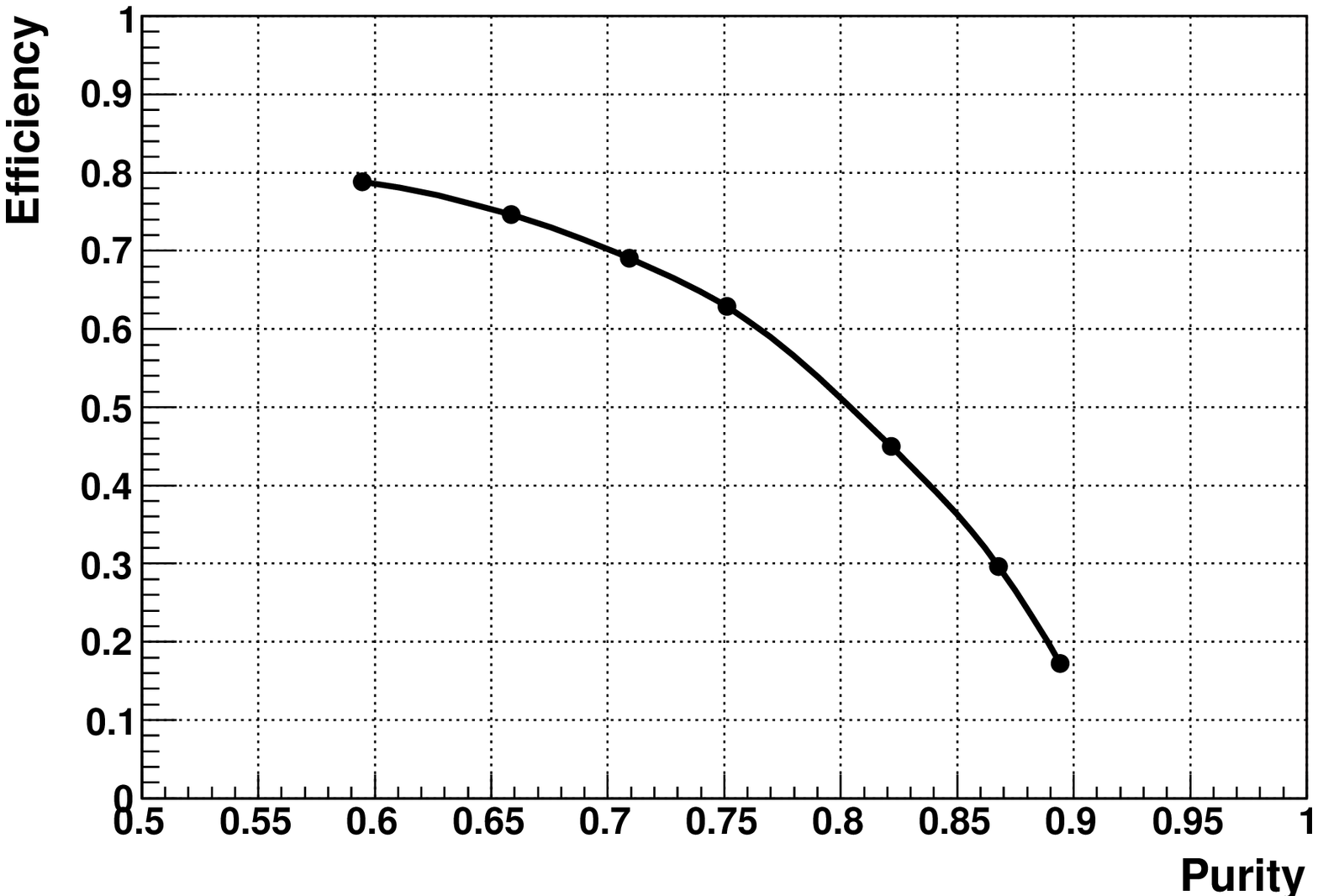}
\caption{The efficiency versus purity for good STT
 tracks relative to good tracks found by the offline reconstruction in a 
 sample dominated by multijet events.  Each point corresponds to a different
 requirement on the fit $\chi^2$ used to define good STT tracks.\label{f-evp}}
\end{figure}

Fig.~\ref{f-zeffi} shows the efficiency measured using collider data for
reconstructing an STT track measured relative to tracks found by the offline
reconstruction program for muon tracks arising from the decay
$Z\rightarrow\mu\mu$.  The horizontal axis is the maximum fit $\chi^2$ for
defining a good track.  A tight requirement corresponds to low $\chi^2$ and
thus lower efficiency.  Finally, Fig.~\ref{f-evp} shows the efficiency as a
function of purity for good STT tracks measured relative to tracks found by the
offline reconstruction program.  Here purity is defined as the fraction of good
STT tracks which match well to a track found by the offline reconstruction
program.


\section{Summary}
The Silicon Track Trigger described in this paper consists of custom hardware
used to identify $\ppbar$ collisions which result in production of
$b$--flavored particles in near real--time.  It resides in the second level of
the D\O\ trigger system and uses inputs from the first level track trigger and
the silicon microstrip detector to provide precision reconstruction of
charged particle trajectories in the $r\phi$ plane.  This system can
significantly enhance the D\O\ physics capability in such diverse areas as
$b$--flavored jet energy calibration, the top mass precision and sensitivity to
searches for the Higgs boson.  The STT fits tracks using a combination of SMT
and CFT information and is particularly geared toward measuring impact
parameters.  The intrinsic precision is roughly 20~$\mu$m for high-$p_T$
tracks, with an additional contribution of roughly 35~$\mu$m from the beam
spot.  The single--hit resolution and multiple scattering effects are of
similar size for tracks of $p_T \approx 2.5$~GeV.  The STT computations
introduce negligible dead time to the D\O\ data taking.  The STT has been in
routine operation since mid 2004.

\newpage
\noindent{\Large\bf Acknowledgments}
\vskip 2mm

The authors would like to thank their D\O\ collaborators for useful discussions
and help with the STT project.  We
thank Florida State University faculty Reginald Perry and his students Shweta
Lolage and Vindi Lalam for their early contributions to STC design and firmware
development. We acknowledge major funding of the project by the National
Science Foundation Major Research Instrumentation program under grant
PHY-997659 and by the U.S. Department of Energy under grant
DE-FG02-91ER40676. We also thank Boston University, Columbia University,
Florida State University, IN2P3 (France), SUNY Stony Brook, FOM-Institute
NIKHEF/University of Nijmegen (The Netherlands) for additional
funding. We gratefully acknowledge in kind contributions from the
Altera and Xilinx corporations. One of us (U.H.) would like to acknowledge
support by the Alfred P. Sloan Foundation and by Research Corporation.


\end{document}